\documentclass[twocolumn]{aastex6}  

\usepackage{grffile}  
\usepackage[all]{hypcap}  
\usepackage[normalem]{ulem}  
\usepackage{multirow, textcomp, verbatim, booktabs}

\newcommand{\foundation}{\textit{foundation}}
\newcommand{\uniform}{\textit{uniform}}
\newcommand{\narchan}{\textit{narchan}}
\newcommand{\widechan}{\textit{widechan}}

\newcommand{\nee}{NEext}
\newcommand{\nea}{NEa}
\newcommand{\neb}{NEb}
\newcommand{\nec}{NEc}
\newcommand{\ned}{NEd}

\newcommand{\hi}{H\textsc{i}}
\newcommand{\subhi}{{\rm H\textsc{\scriptsize i}}}

\newcommand{\fdnrms}{1.9}  
\newcommand{\fdncolumneight}{1.4}  
\newcommand{\nchrms}{2.0}  
\newcommand{\nchcolumneight}{0.8}  
\newcommand{\wchrms}{1.6}  
\newcommand{\wchcolumneight}{2.4}  
\newcommand{\momzmin}{18.7}  
\newcommand{\incl}{44}  
\newcommand{\eincl}{6}  
\newcommand{\vsyshi}{235.7}  
\newcommand{\evsyshi}{0.6}  
\newcommand{\wihi}{38.7}  
\newcommand{\ewihi}{4.8}  
\newcommand{\hiflux}{1.1}  
\newcommand{\ehiflux}{0.1}  
\newcommand{\mhigalsix}{8.4}  
\newcommand{\emhigalsix}{1.1}  
\newcommand{\mhitotsix}{8.9}  
\newcommand{\emhitotsix}{1.1}  
\newcommand{\mhineabfive}{4.5}  
\newcommand{\mhineafive}{2.1}  
\newcommand{\emhineafive}{1.4}  
\newcommand{\mhinebfive}{2.6}  
\newcommand{\emhinebfive}{1.6}  
\newcommand{\mhinecfive}{1.5}  
\newcommand{\emhinecfive}{1.2}  
\newcommand{\mhinedfive}{2.4}  
\newcommand{\emhinedfive}{1.4}  
\newcommand{\vdisp}{6.2}  
\newcommand{\evdisphi}{1.3}  
\newcommand{\evdisplo}{1.6}  
\newcommand{\vrot}{17.2}  
\newcommand{\evrot}{7.7}  
\newcommand{\rrot}{1.75}  
\newcommand{\mdyneight}{1.7}  
\newcommand{\emdyneight}{1.1}  
\newcommand{\extrabarymassfrac}{3.5}  
\newcommand{\totalbaryonfraction}{0.13}

\newcommand{\ebaryonfraction}{0.17}  

\shorttitle{\hi\ in Pisces A}
\shortauthors{Beale et al.}

\begin{document}

\title{The \hi\ Structure of the Local Volume Dwarf Galaxy Pisces A}

\author{Luca Beale\altaffilmark{1,2,}$^\dagger$, Jennifer Donovan Meyer\altaffilmark{2}, Erik J. Tollerud\altaffilmark{3}, Mary E. Putman\altaffilmark{4}, and J. E. G. Peek\altaffilmark{3,5}}
\altaffiltext{1}{Department of Astronomy, University of Virginia, Charlottesville, VA 22904, USA}
\altaffiltext{2}{National Radio Astronomy Observatory, Charlottesville, VA 22901, USA}
\altaffiltext{3}{Space Telescope Science Institute, 3700 San Martin Drive, Baltimore, MD 21218, USA}
\altaffiltext{4}{Department of Astronomy, Columbia University, Mail Code 5246, 550 West 120th Street, New York, NY 10027, USA}
\altaffiltext{5}{Department of Physics \& Astronomy, Johns Hopkins University, 3400 North Charles Street, Baltimore, MD 21218, USA}

\email[$^\dagger$]{lb5eu@virginia.edu}

\begin{abstract}
Dedicated \hi\ surveys have recently led to a growing category of low-mass galaxies found in the Local Volume. We present synthesis imaging of one such galaxy, Pisces A, a low-mass dwarf originally confirmed via optical imaging and spectroscopy of neutral hydrogen (\hi) sources in the Galactic Arecibo L-band Feed Array \hi\ (GALFA-\hi) survey. Using \hi\ observations taken with the Karl G. Jansky Very Large Array (JVLA), we characterize the kinematic structure of the gas and connect it to the galaxy's environment and evolutionary history. While the galaxy shows overall ordered rotation, a number of kinematic features indicate a disturbed gas morphology. These features are suggestive of a tumultuous recent history, and represent $\sim \extrabarymassfrac$\% of the total baryonic mass. We find a total baryon fraction $f_{\rm bary} = \totalbaryonfraction$ if we include these features. We also quantify the cosmic environment of Pisces A, finding an apparent alignment of the disturbed gas with nearby, large scale filamentary structure at the edge of the Local Void. We consider several scenarios for the origin of the disturbed gas, including gas stripping via ram pressure or galaxy-galaxy interactions, as well as accretion and ram pressure compression. Though we cannot rule out a past interaction with a companion, our observations best support the suggestion that the neutral gas morphology and recent star formation in Pisces A is a direct result of its interactions with the IGM.
\end{abstract}


\section{Introduction} \label{sec:intro}

\begin{figure*} \label{fig:hst+cont}
\includegraphics[width=\textwidth]{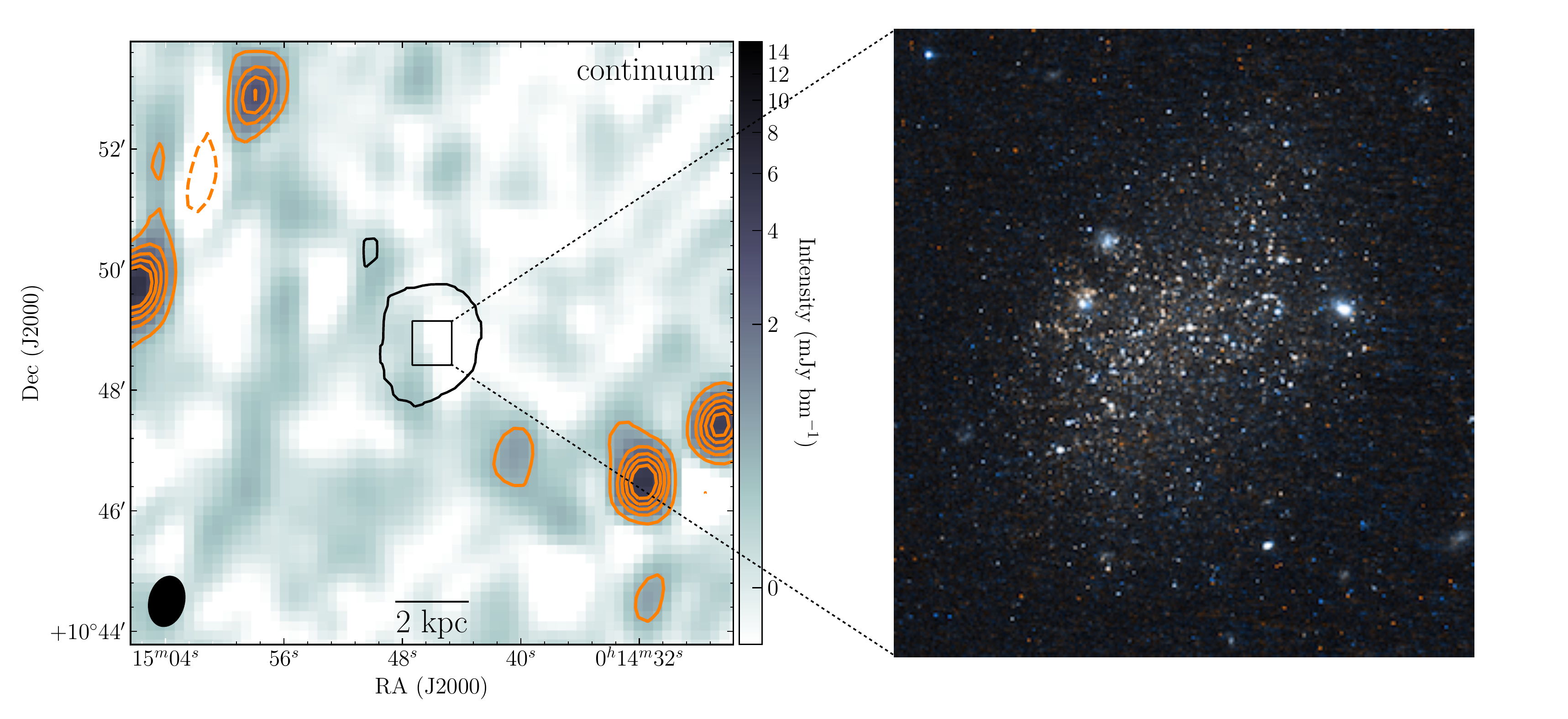}
\caption{
	Continuum emission (derived from 3$\times$128 MHz subbands centered on 1435.5, 1563.5, and 1691.5 MHz) in a $5\arcmin \times 5\arcmin$ region around Pisces A. Orange contours highlight continuum sources at the depth of our imaging. The black \momzmin~mJy bm$^{-1} \cdot$ km s$^{-1}$ contour highlighting the extent of the \hi\ emission is from the \foundation-derived integrated intensity map (see Section \ref{sub:cubes}). The $44\arcsec \times 40\arcsec$ inset color image shows the optical extent of the galaxy. The colors approximate the human eye response to the F606W/F814W \textit{HST} ACS imaging \citep[for details, see][]{tollerud+2016}.
}
\end{figure*}

Modulo internal processes, the survival of gas in galaxies throughout cosmic history generally suggests that gas is either being replenished (via mergers or accretion) or shielded by a halo. In cluster environments, such shielding can happen in large dark matter subhaloes \citep{penny+2009}, which prevent disruption from the cluster potential. Gas may be accreted at late times from the cosmic web \citep{keres+2005}, and \citet{ricotti2009} suggests that accretion from the intergalactic medium may still occur even after reionization. Understanding these processes is crucial to determining the overall growth of dwarf galaxies, and how their star formation and interstellar medium (ISM) are coupled \citep{dekel+1986, kennicutt1998, brooks+2014}. Despite low column densities \citep[$\lesssim 10^{21}\rm\ cm^{-2}$ integrated over a $\sim20\arcsec$ beam, corresponding to $\sim300-700$ pc for the SHIELD sample;][]{cannon+2011}, dwarfs can still have ongoing star formation at low masses ($M_\subhi \sim 10^{6-7}\rm\ M_\odot$). The mechanisms by which dwarfs consume their fuel (and the timescales involved) are complex and poorly studied, especially when \hi\ persists in extended structures around these galaxies \citep{roychowdhury+2011}.

The cosmic environment of dwarf galaxies can also provide insights into the evolution of their gas content. Low-mass galaxies are thought to accrete their gas mainly through the so-called ``cold mode'', even down to $z = 0$ \citep{keres+2005, keres+2009, maddox+2015}. The low temperature ($T \lesssim 10^5$ K) gas is expected to be transported along cosmic filaments, allowing galaxies to funnel gas from large distances. Possible evidence for such accretion may be observable in the gas kinematics of individual galaxies, e.g., the misalignment of the kinematical major and minor axes \citep[KK 246;][]{kreckel+2011b}, warping of the polar disk \citep[J102819.24+623502.6;][]{stanonik+2009}, and disturbed morphology \citep[KKH 86;][]{ott+2012}. This cold mode accretion mechanism is also expected to be more significant in low density regions, and a handful of individual systems from the Void Galaxy Survey \citep[VGS;][]{kreckel+2014}, in which the morphology and kinematics of 59 galaxies selected to reside within the deepest voids identified in the Sloan Digital Sky Survey (SDSS) were studied, reveal kinematic warps and low column density envelopes consistent with ``cold'' gas accretion \citep[see especially Figures 9 and 10 in][]{kreckel+2012}.

Here we present resolved \hi\ imaging of one such low-mass dwarf galaxy taken with the Karl G. Jansky Very Large Array\footnote{The National Radio Astronomy Observatory is a facility of the National Science Foundation operated under cooperative agreement by Associated Universities, Inc.} (JVLA). Pisces A, a Local Volume dwarf galaxy, was originally identified in the GALFA-\hi\ DR1 catalog \citep{peek+2011} as an \hi\ cloud with a size $< 20'$ and a velocity FWHM $< 35 \rm \ km\ s^{-1}$ \citep{saul+2012, tollerud+2015}. Followup optical spectroscopy \citep{tollerud+2016} confirmed the coincidence of the \hi\ emission with a stellar component, cementing its status as a galaxy. \citet{carignan+2016} obtained interferometric observations with the compact Karoo Array Telescope (KAT-7). With a $5.3\arcmin \times 3.4\arcmin$ synthesized beam (natural weighting), they reported warping of the \hi\ disk toward the receding side.

From the location of Pisces A within the Local Volume, and its star formation history (SFH), \citet{tollerud+2016} suggest that the galaxy has fallen into a higher-density region from a void. If this is indeed the case, subsequent gas accretion or encounters with other galaxies could trigger delayed (recent) evolution and leave an imprint on the gas content observable through the kinematics and morphology. We explore this possibility here by examining the resolved gas kinematics of Pisces A.

This paper is organized as follows: In Section \ref{sec:data}, we describe the JVLA observations. Our techniques for extracting kinematic and morphological information are presented in Section \ref{sec:methods}. We highlight the velocity structure and kinematics in Section \ref{sec:results}. We place Pisces A in context in Section \ref{sec:context}, discuss possible origins of the disturbed gas in Section \ref{sec:discussion} and conclude in Section \ref{sec:conclusions}.

\section{Observations and Data Reduction} \label{sec:data}

Pisces A was observed in December 2015 with the JVLA in D-configuration for 2h, corresponding to $\sim1.3$h of on-source integration time (JVLA/15B-309; PI Donovan Meyer). The WIDAR correlator was configured to provide 4 continuum spectral windows (128 MHz bandwidth subdivided into 128 channels) and a single spectral window centered on the \hi\ line (8 MHz subdivided into 2048 channels). After Hanning smoothing and RFI excision, we perform standard calibration of the visibility data within the Common Astronomy Software Applications \citep[CASA;][]{mcmullin+2007} environment. J0020+1540 is used to calibrate the phase and amplitude, while 3C48 serves as a bandpass and flux calibrator. The calibrated data are then split off into their own MeasurementSet. Baselines with noisy or bad data are removed before further analysis.

\autoref{tab:properties} summarizes the parameters of this galaxy, and an optical image is shown in \autoref{fig:hst+cont}.

\section{Methods} \label{sec:methods}

\subsection{Data Cubes} \label{sub:cubes}

\begin{deluxetable}{lccc}  
\tabletypesize{\footnotesize}
\tablecaption{Key Parameters of Pisces A \label{tab:properties}}
\tablehead{
Parameter								&	\colhead{Value}					&	\colhead{Reference}
}
\startdata 
RA (J2000)							&	$\rm 00^h14^m46^s$			&	\\
Dec. (J2000)							&	$+10^\circ48\arcmin47.01\arcsec$	&	\\
Distance (Mpc)							&	$5.64^{+0.15}_{-0.13}$			&	(3) \\
Optical axial ratio ($q=b/a$)				&	$0.66 \pm 0.01$				&	(3) \\
Optical inclination$^a$ ($^\circ$)			&	$59 \pm 3$					&	(1, 4) \\
\hi\ axial ratio						&	$0.79 \pm 0.02$				&	(1) \\
\hi\ inclination$^{a}$ ($^\circ$)				&	$\incl \pm \eincl$				&	(1) \\
$M_V$ (mag)							&	$-11.57^{+0.06}_{-0.05}$			&	(3) \\
$r_{\rm eff, major}$ (pc)					&	$145^{+5}_{-6}$				&	(3) \\
$\log(M_\star/\rm M_\odot)$				&	$7.0^{+0.4}_{-1.7}$				&	(3) \\
$v_{\rm sys, opt}$ (km s$^{-1}$)			&	$240 \pm 34$					&	(2) \\
\hline
\multicolumn{3}{c}{Arecibo} \\
\hline
$v_{\rm sys, \subhi}$ (km s$^{-1}$)			&	$236 \pm 0.5$					&	(2) \\
$W50_\subhi$ (km s$^{-1}$)				&	$22.5 \pm 1.3$					&	(2) \\
$M_\subhi$ ($10^6\rm\ M_\odot$)			&	$8.9 \pm 0.8$					&	(2, 3) \\
\hline
\multicolumn{3}{c}{KAT-7 Array} \\
\hline
$v_{\rm sys, \subhi}$ (km s$^{-1}$)			&	$233 \pm 0.5$					&	(4) \\
$W50_\subhi$ (km s$^{-1}$)				&	$28 \pm 3$					&	(4) \\
$M_\subhi$ ($10^6\rm\ M_\odot$)			&	$13 \pm 4$					&	(3, 4) \\
\hline
\multicolumn{3}{c}{JVLA} \\
\hline
$v_{\rm sys, \subhi}$(km s$^{-1}$)			&	$\vsyshi \pm \evsyshi$			&	(1) \\
$W50^i_\subhi$ (km s$^{-1}$)				&	$\wihi \pm \ewihi$				&	(1) \\
$M_{\subhi,\rm gal}$ ($10^6\rm\ M_\odot$)	&	$\mhigalsix \pm \emhigalsix$		&	(1) \\
\enddata
\tablenotetext{}{(1) This work.}
\tablenotetext{}{(2) \citet{tollerud+2015}}
\tablenotetext{}{(3) \citet{tollerud+2016}}
\tablenotetext{}{(4) \citet{carignan+2016}}
\tablenotetext{}{$^a$Assuming $q_0 = 0.48 \pm 0.04$ \citep{roychowdhury+2013}.}
\end{deluxetable}

When imaging the \hi\ spectral window, we restrict the data to a spectral range of 100 km s$^{-1}$ centered on the systemic velocity, as we find no significant emission beyond this range. We apply a primary beam correction to all data cubes. Note that none of our final data cubes are continuum-subtracted. Although continuum sources exist in the field at the depth of our imaging, no significant continuum emission is coincident with Pisces A (left panel of \autoref{fig:hst+cont}).

Before creating cubes for analysis, we image subsets of the data to affirm the detection of the features discussed in Section \ref{sec:results}. In particular, we split and image the calibrated visibilities by polarization (RR and LL) as well as by scan (`first half' and `second half' of the night) and analyze them separately. This reveals a handful of bad baselines affecting the `first half' scans, which we subsequently remove. However, broadly speaking, we find little difference in the noise characteristics of any images created using the data subsets, and our results are not significantly changed by flagging the affected baselines.

To maximize our sensitivity to a range of spatial scales, and to explore whether the features detected at one spatial/velocity resolution might also be apparent at another spatial/velocity resolution, we produce multiple versions of the data cube with different imaging parameters. We create cubes with three spatial weighting functions (uniform, Briggs weighting, and natural) and 2 km s$^{-1}$ channels, as well as two additional Briggs weighted cubes with narrower (1 km s$^{-1}$) and wider (4 km s$^{-1}$) channels. The three cubes used in our final analysis are summarized in \autoref{tab:cubes}. 

Among our three cubes with varying spatial resolution, we find that Briggs weighting \citep[with robust = 0.5;][]{briggs1995} yields similar results to natural weighting in terms of rms noise and the spectra derived at the locations of our detected features, with a slightly smaller beam size. Thus we present analyses based on the robust weighted cube (\foundation) assuming that the beam size is a better match to the linear size of the detected features discussed in Section \ref{sec:results}. The \uniform\ data cube, on the other hand, provides increased spatial resolution to capture more compact emission; in the end, this cube did not reveal new information about Pisces A and is not highlighted in the paper.

Improved spectral resolution in the \narchan\ data cube is achieved by binning to 1 km s$^{-1}$ velocity resolution. While this results in decreased signal-to-noise, the smaller channel width allows us to distinguish between kinematic components, and offers a better estimate of the rotation curve. Finally, the \widechan\ data cube is binned to 4 km s$^{-1}$ channels for increased sensitivity to extended structure, at the cost of spectral resolution. 

\begin{deluxetable*}{lccccccc}  
\tablecaption{Properties of the Data Cubes\label{tab:cubes}}
\tablehead{
Cube		&	\colhead{$\Delta v$}	&	\colhead{$b_{\rm maj}$}	&	\colhead{$b_{\rm min}$}	&	\colhead{$\theta_{\rm PA}$}	&	\colhead{$s_{\rm pix}$}	&	\multicolumn{2}{c}{$\sigma_{\rm rms}$
} \\
\cmidrule(lr){7-8}  
			&	(km s$^{-1}$)		&	(\arcsec)				&	(\arcsec)				&	($^\circ$)					&	(\arcsec/pix)			&	(mJy bm$^{-1}$)	&	($10^{18}$~atoms~cm$^{-2}$)  
} 
\startdata 
\foundation	& 	2 				& 	64.08 				&	45.74				&	$-11.67$					&	8					&	\fdnrms			&	\fdncolumneight \\
\narchan 		& 	1 				& 	64.08				&	45.75				&	$-11.77$					&	8					&	\nchrms			&	\nchcolumneight \\
\widechan		& 	4 				& 	64.00				&	45.80				&	$-11.66$					&	8					&	\wchrms			&	\wchcolumneight \\
\enddata
\end{deluxetable*}

\begin{figure*}
\includegraphics[width=\textwidth]{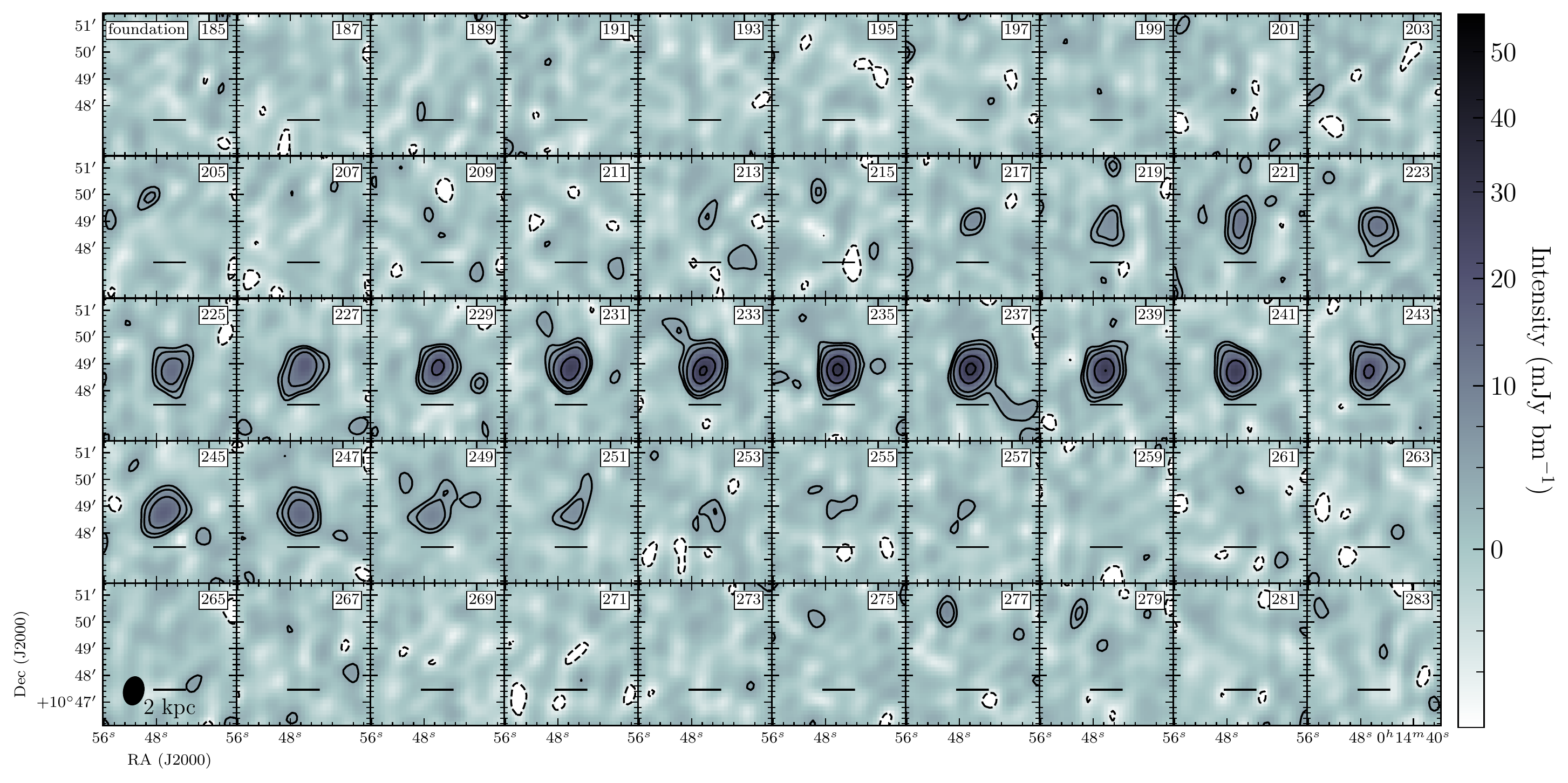}
\caption{\small Channel maps from the \foundation\ data cube of Pisces A. Contour levels are at ($-$2, 2, 3, 5, 10, 15)$\sigma_{\rm rms}$ where $\sigma_{\rm rms} = \fdnrms\rm\ mJy\ bm^{-1}$. The velocity (in km s$^{-1}$) is marked in the upper right corner of each panel. The JVLA beam is given in the lower left panel.}
\label{fig:chanmap}
\end{figure*}

In order to construct moment maps for Pisces A, we first apply a spatial mask to the data cube, derived from the shape of the emission at 233 km s$^{-1}$ in the \foundation\ data cube (chosen to encompass the northeast extension; see \autoref{fig:chanmap} and Section \ref{sub:north}). Then, we mask out any emission below the $3\sigma_{\rm rms}$ level. Finally, for the pixels that remain, we mask out any that are spectrally isolated -- that is, a pixel is masked out if neither of its spectral neighbors have remained. We then produce moment 0 (integrated intensity), moment 1 (intensity-weighted velocity), and moment 2 (intensity-weighted velocity dispersion) maps from the masked data cube. Although physical interpretation of moment 1 and moment 2 maps as ``velocity fields'' and ``dispersion maps'', respectively, is valid only for a true Gaussian line profile, we use these terms interchangeably for simplicity. Global \hi\ profiles are constructed by averaging over the same spatial mask. All velocity information is presented in the kinematic Local Standard of Rest (LSRK) frame\footnote{Both \citet{tollerud+2016} and \citet{carignan+2016} report velocities in the heliocentric reference frame. Assuming no proper motion, the LSRK frame is $\lesssim 0.5\rm\ km\ s^{-1}$ offset from the heliocentric frame at the location of Pisces A.} unless otherwise noted.

The inclination of the galaxy is calculated by \citep{holmberg1958}
\begin{equation}\label{eqn:sini}
\sin^2i = \frac{1 - q^2}{1 - q_0^2},
\end{equation}
where $q = b/a$ is the observed axial ratio and $q_0$ is the intrinsic axial ratio. The former is calculated from the \foundation-derived moment 0 map, with uncertainties estimated by a Monte Carlo simulation (see \autoref{tab:properties}). For the latter, we adopt $q_0 = 0.48 \pm 0.04$, corresponding to the peak of the distribution found by \citet{roychowdhury+2013} for faint dIrrs in the Local Volume. To capture the impact of the standard deviation $\sigma_{q_0}$ on the inclination, we perform a Monte Carlo simulation by randomly sampling the distribution of $q_0$ and measuring the standard deviation of the resulting distribution of inclination angles.  The final error on the inclination angle is then the quadrature sum of the standard deviation with systematic errors.

\subsection{PV Diagrams}\label{sub:pvdiagrams}

Position-velocity (PV) diagrams are constructed by extracting velocity information along a pre-defined axis chosen to encompass the total spatial extent of the emission. An approximate ``major'' axis is defined using the \hi\ isovelocity contours, and an ``extension'' axis is defined to cross the northeast extension (see Section \ref{sub:north} and \autoref{fig:pv}). Both axes are centered on the coordinates of Pisces A as given in \autoref{tab:properties} so that a $0\arcsec$ offset corresponds to the optical center of the galaxy.

\subsection{Rotation Curves}\label{sub:rotcur}

Well-constrained analyses of rotation curves using \hi\ observations require both high spectral and spatial resolution, such that multiple beams can be placed across the disk of the galaxy. Commonly, this is done by assuming the velocity field can be fit by a series of tilted rings (whose width is set by the beam size), allowing one to extract a detailed rotation curve \citep[e.g., \texttt{$^{\tt 3D}$BAROLO};][]{diteodoro+2015}. The extent of the \hi\ in Pisces A ($\sim 2'$ across, corresponding to 2-3x the beam size) prohibits us from taking advantage of this method\footnote{\texttt{$^{\tt 3D}$BAROLO} is built to handle barely resolved galaxies. However, in this case, the number of (unknown) free parameters involved in the 3D fit makes interpretation of the results difficult.}. Instead, we use a variation of the peak intensity method \citep[][see also \citeauthor{sofue+2001} \citeyear{sofue+2001} and \citeauthor{takamiya+2002} \citeyear{takamiya+2002}]{mathewson+1992} to estimate the rotation curve.

\begin{figure*}
\gridline{
\fig{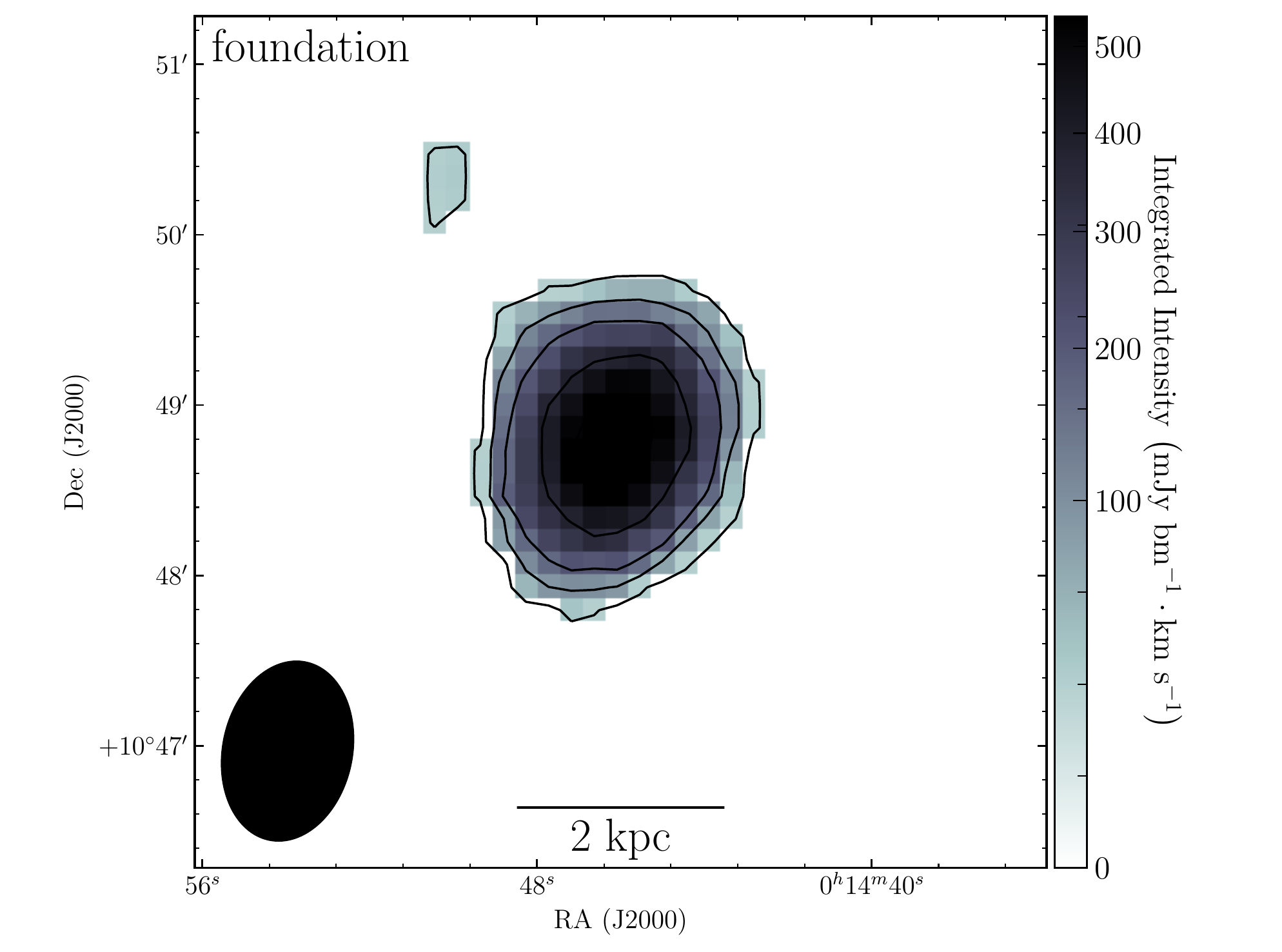}{0.48\textwidth}{\vspace{-6.5cm}\hspace{4.8cm}\textbf{\large a}}
\fig{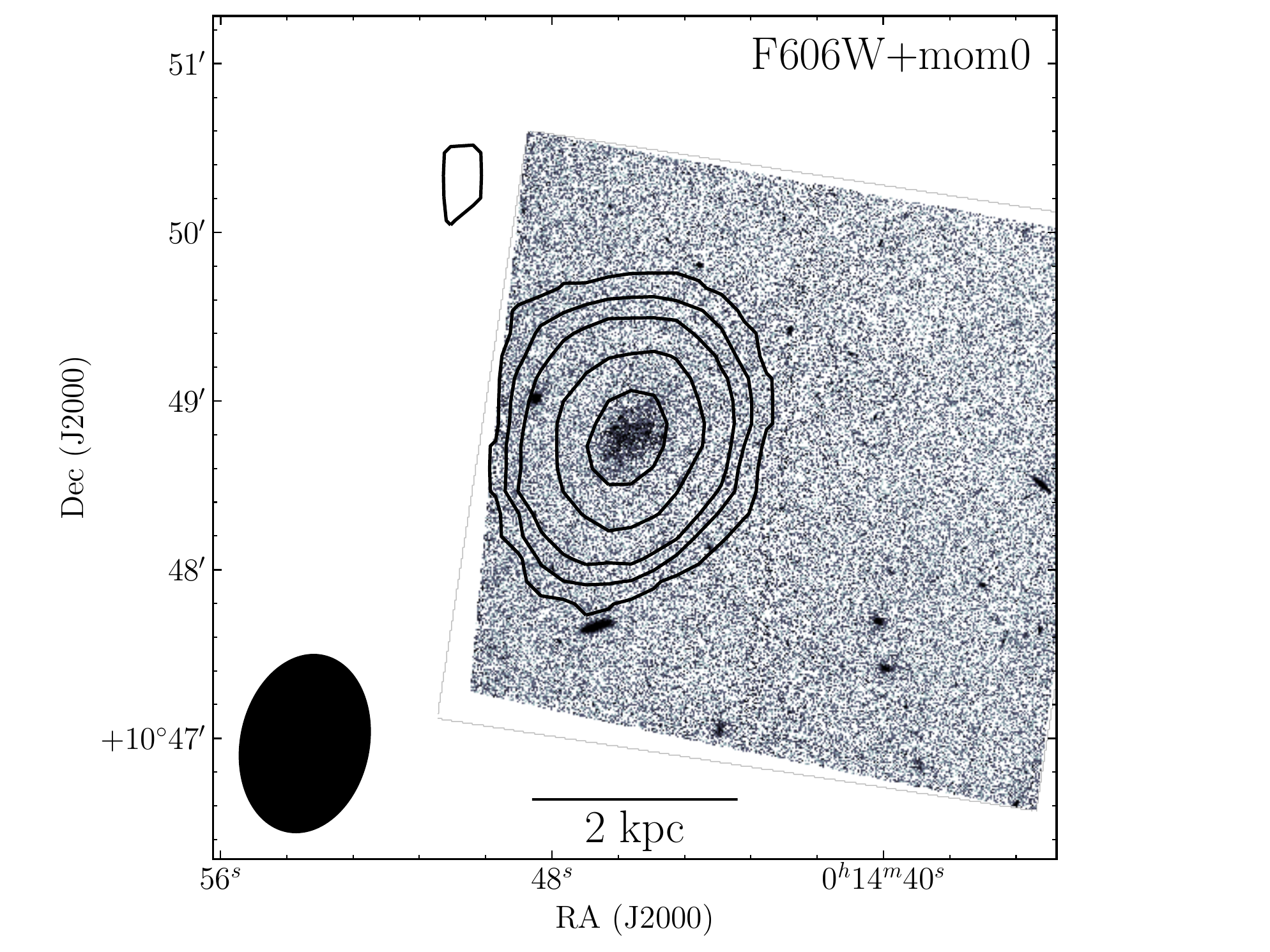}{0.48\textwidth}{\vspace{-6.6cm}\hspace{-4.9cm}\textbf{\large b}}
}
\vspace{-0.5cm}
\gridline{
\vspace{-0.5cm}
\fig{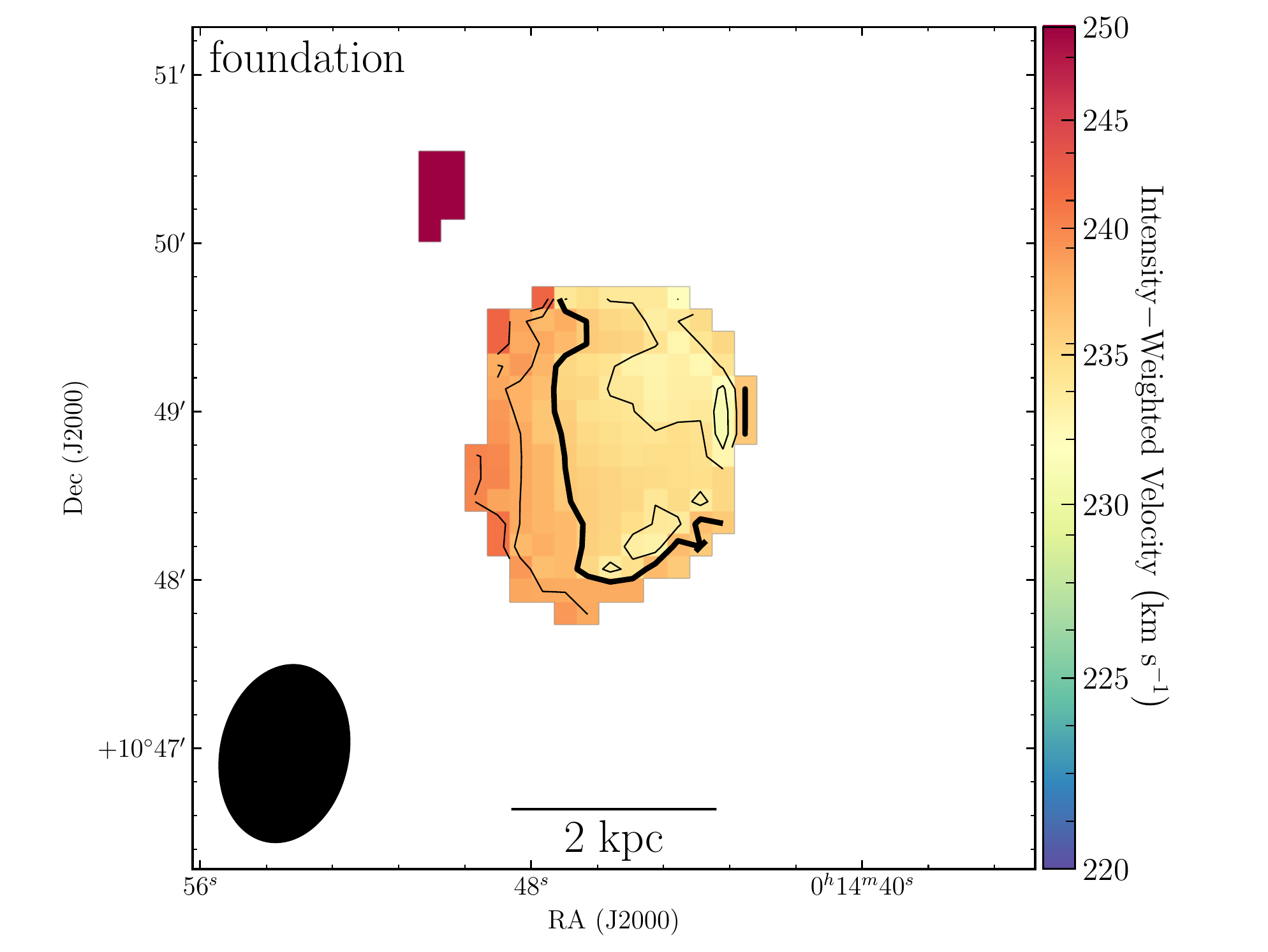}{0.48\textwidth}{\vspace{-6.5cm}\hspace{4.8cm}\textbf{\large c}}
\fig{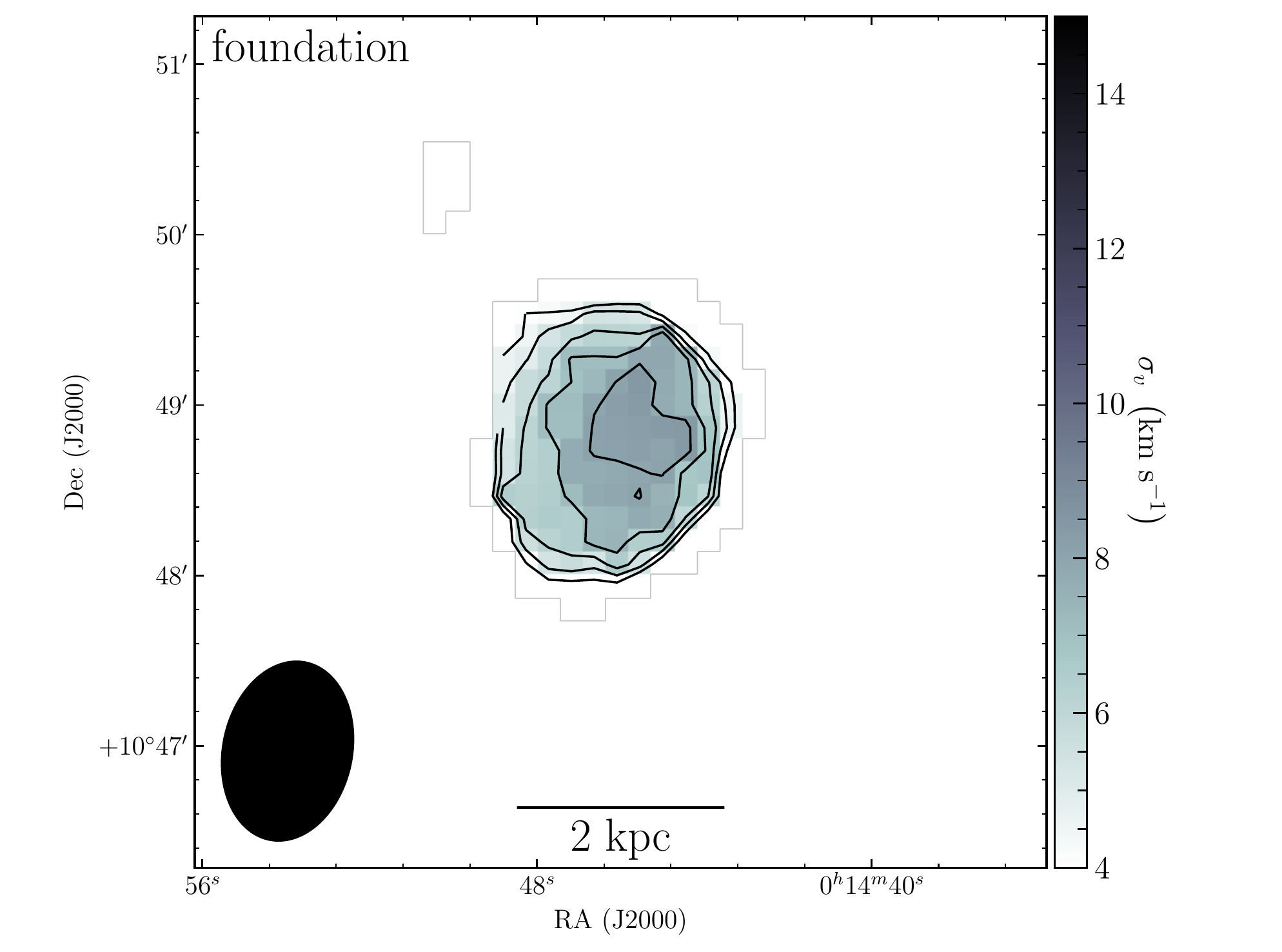}{0.48\textwidth}{\vspace{-6.5cm}\hspace{4.8cm}\textbf{\large d}}
}
\caption{\small Moment analysis of the \foundation\ data cube. The JVLA beam is given in the bottom left corner of each panel. All moment maps are constructed according to Section \ref{sub:cubes}. (a) Integrated intensity map. Contour levels are at (1, 5, 10, 20, 30)$\times 18.7$ mJy bm$^{-1} \cdot$ km s$^{-1}$. (b) \textit{HST} ACS F606W imaging of Pisces A. Contours are the same as in the top left. Background sources can be seen toward the southwest of the field. (c) Velocity field of the galaxy. Isovelocity contours are spaced every 2 km s$^{-1}$, and the thick black contour denotes the systemic velocity. (d) Dispersion map with contours spaced every 2 km s$^{-1}$.}
\label{fig:moment}
\end{figure*}

From the major-axis PV diagram, we select the pixels with maximum intensity across the spatial direction (i.e., sampling at each spatial pixel, thus oversampling the beam by a factor of $\sim$5). After masking to emission at the $\geq 2\sigma_{\rm rms}$ level and removing outlier bright pixels believed not to follow the galaxy rotation, we rebin the data to only sample every half-beam. To estimate the associated uncertainties, we fit a Gaussian to the intensities at each pixel bin, and treat the resulting estimate of $v_{\rm rot}$ as the ``true'' solution. We then draw random samples from a distribution with the same parameters as the ``true'' solution and re-fit, repeating this 1000 times. The errors are then the magnitude of the median difference between the individual fits and the ``true'' solution, i.e., $\left| {\rm median}\left(v_{\rm rot, fit} - v_{\rm rot, true}\right)\right|$, added in quadrature with the channel width for that cube. All rotation velocities are corrected for inclination.

\section{Results}\label{sec:results}

\subsection{Overall \hi\ Content and Kinematics}\label{sub:overall}

Channel maps for the \foundation\ cube are presented in \autoref{fig:chanmap}. In general, the channel maps reveal regular emission across the main body of the galaxy, which is also seen in other data cubes. However, multiple features are observed which are inconsistent with the general rotation, which we discuss in detail in Section \ref{sub:north}.

In \autoref{fig:moment} we show the moment analysis for Pisces A, as applied to the \foundation\ data cube. The overall \hi\ distribution (top left panel) is centered on the optical component of the galaxy and reveals a `clumpy' feature toward the northeast. This is also seen in the channel maps at multiple velocities. Another feature toward the southwest appears in the channel maps, but does not `survive' the moment construction. The top right panel of \autoref{fig:moment} reveals multiple background galaxies that are spatially coincident, but obviously not associated, with this emission. Both of these features are discussed further in Section \ref{sub:north}.

The velocity field of Pisces A is shown in the bottom left panel of \autoref{fig:moment}. Pisces A exhibits approximately solid-body rotation out to the extent of the \hi, consistent with typical expectations of dwarf galaxies. There is a slight asymmetry on the approaching side, which is approximately at the edge of the optical disk, indicating a possible warp. The dispersion map in the bottom right panel suggests an approximately rotation-dominated body, though we note that the area of highest velocity dispersion is not apparently centered on the stars.

\begin{figure}
\includegraphics[width=\linewidth]{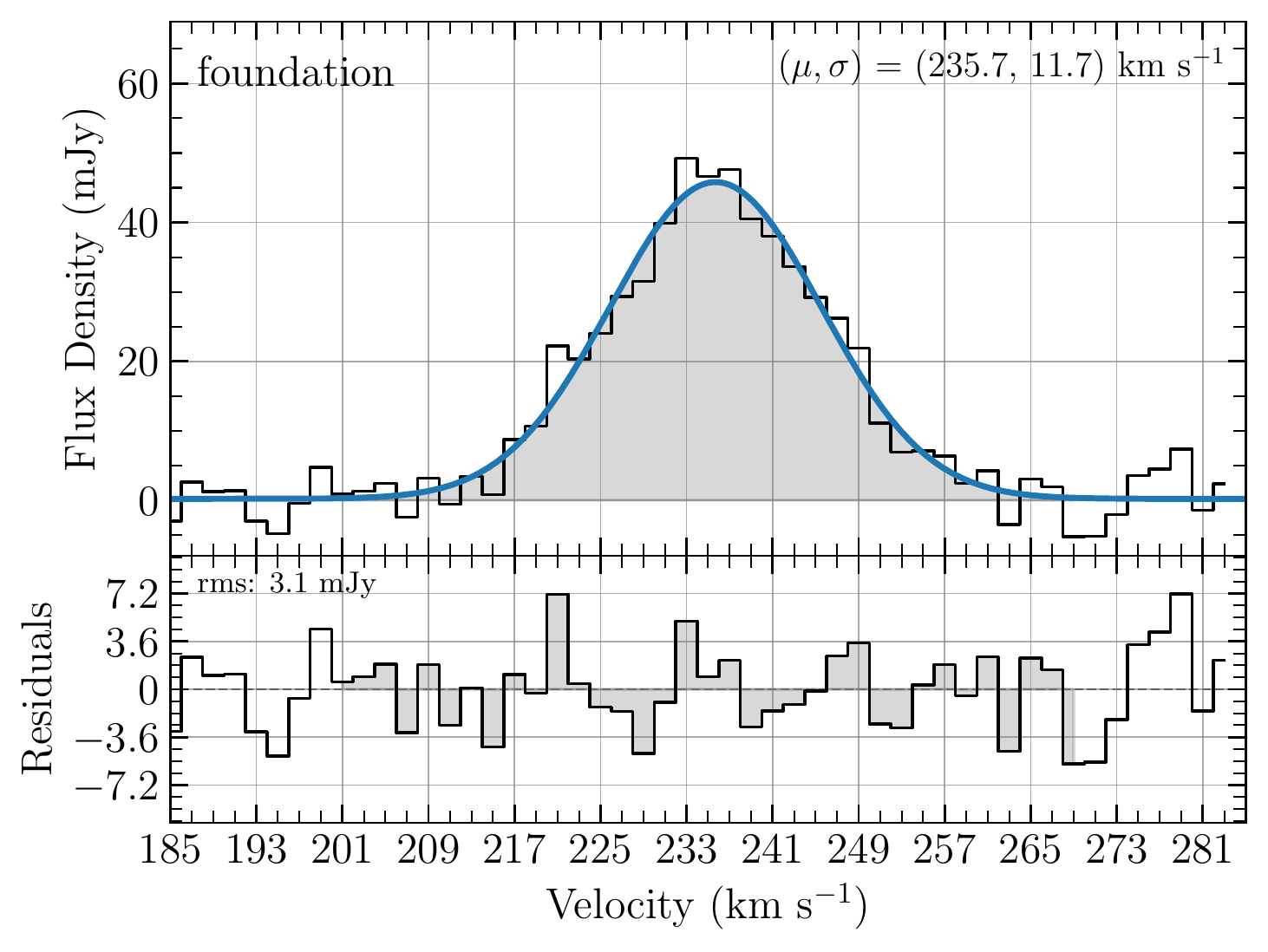}
\caption{\small Global \hi\ spectrum extracted from the \foundation\ cube. The blue solid curve is the best fit Gaussian. In both panels, the shaded region (containing $\pm 3\sigma =$ 99.7\% of the emission) denotes the range of emission used in further calculations.}
\label{fig:spectrum}
\end{figure}

The global \hi\ spectrum of the galaxy is presented in \autoref{fig:spectrum}. We find a systemic velocity of $v_{\rm sys, \subhi} = \vsyshi \pm \evsyshi$ km s$^{-1}$, in agreement with previous single dish and compact array (i.e., KAT-7) observations. The (inclination corrected) 50\% profile width is $W50^i_\subhi = \wihi \pm \ewihi$ km s$^{-1}$. We also find a total \hi\ flux of $F_\subhi = \hiflux \pm \ehiflux$ Jy km s$^{-1}$, corresponding to an \hi\ mass of $M_{\subhi,\rm gal} = (\mhigalsix \pm \emhigalsix) \times 10^6\rm\ M_\odot$. Note that this mass includes contributions from the non-rotating features discussed below, namely the \neb\ and \nec\ components of the northeast extension (see Section \ref{sub:north}). The other features that are not included, \nea\ and \ned, contribute $\mhineabfive \times 10^5\rm\ M_\odot$, giving a total \hi\ mass of $M_{\subhi,\rm tot} = (\mhitotsix \pm \emhitotsix) \times 10^6\rm\ M_\odot$.

\begin{figure*}
\gridline{\vspace{-0.5cm}\hspace{-0.5cm}
\fig{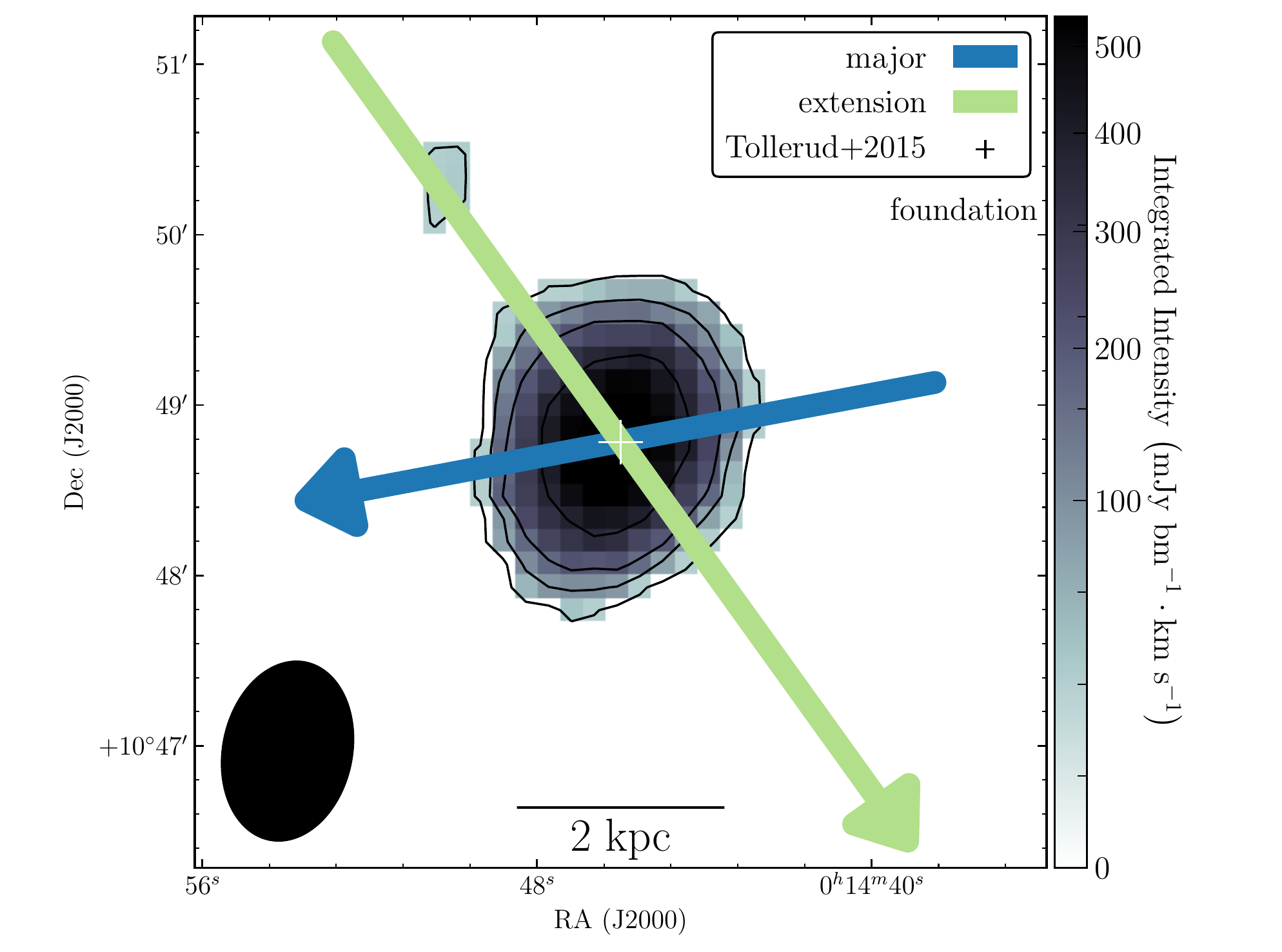}{0.35\linewidth}{}\hspace{-0.5cm}
\fig{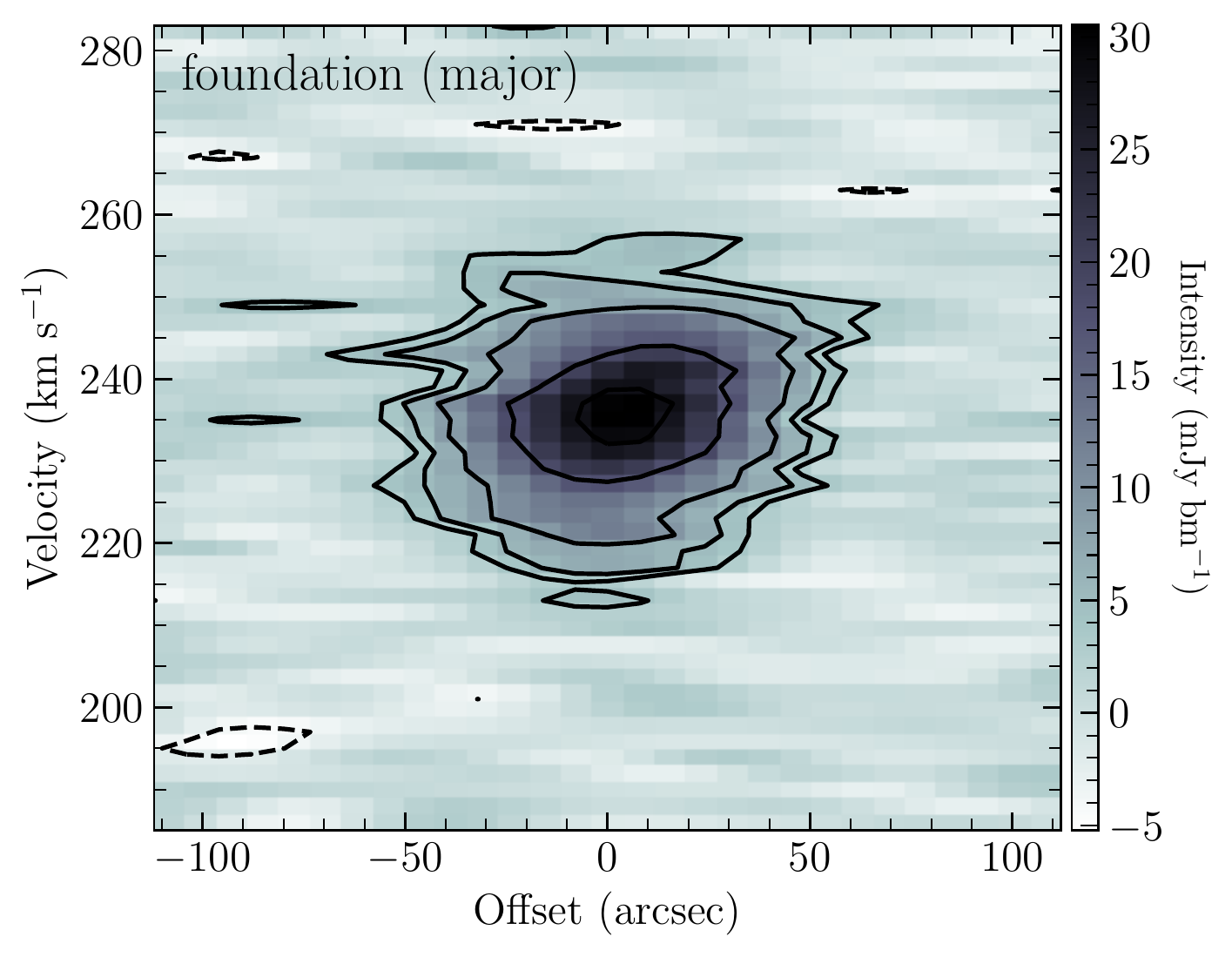}{0.35\linewidth}{}
\fig{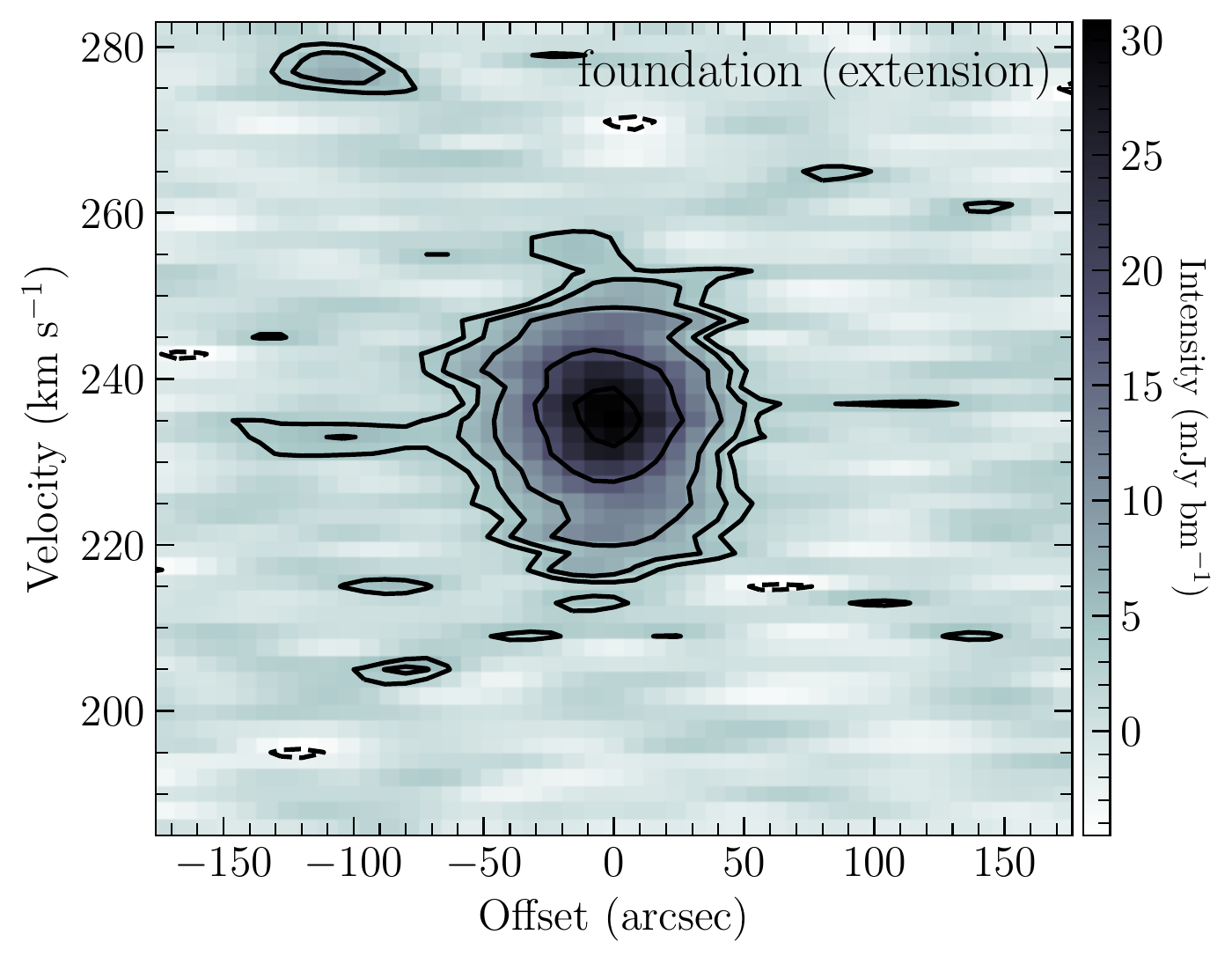}{0.35\linewidth}{}
}
\caption{\small \textit{Left}: Moment 0 map of Pisces A demonstrating the construction of the position velocity slices. Integrated intensity contours are the same as in \autoref{fig:moment}. The optical center of the galaxy is denoted by the cross. Arrows denote the direction of increasing offset. \textit{Center} and \textit{Right}: Corresponding major- and extension-axis PV diagrams. Contours are at ($-$2, 2, 3, 5, 10, 15)$\sigma_{\rm rms}$, where $\sigma_{\rm rms} = \fdnrms \rm\ mJy\ bm^{-1}$. The northeast extension is approximately near $-100\arcsec$ in the extension-axis PV diagram.}
\label{fig:pv}
\end{figure*}

PV diagrams for Pisces A are shown in \autoref{fig:pv}, along with an integrated intensity map demonstrating the selection process. The overall rotation appears shallow, with the faintest contours contributing little to the bulk motion. In both the major- and extension-axis PV diagrams, we observe fragmented emission at multiple velocities and spatial offsets, which we next discuss in turn.

\subsection{Beyond Bulk Motion: The Northeast Extension}\label{sub:north}

One of the most striking features we detect is a `filament' of gas extending toward the northeast. We call this the ``northeast extension'' (\nee) when referring to \textit{all} spatially coincident emission in this region. In \autoref{fig:chanmap}, it is most obvious across a few channels near 233 km s$^{-1}$, but it also appears in the moment maps (\autoref{fig:moment}), and the extension-axis PV diagram in \autoref{fig:pv} suggests that there are multiple kinematically distinct features in that region.

To assess the significance of the brightest components, we extract a global spectrum of the \nee\ which we show in the right panels of \autoref{fig:extension}. We observe four kinematically distinct components at (approximately) 205, 233, 248, and 277 km s$^{-1}$, which we label as \nea, \neb, \nec, and \ned, respectively. In all data cubes, these features are detected at the $\sim 3\sigma_{\rm rms}$ level.

We determine the significance of each feature directly from its spatially integrated spectrum to avoid beam smearing effects. After subtracting off the best fit Gaussians, we compute the noise as the RMS of the residuals (bottom spectra in the right panels of \autoref{fig:extension}). We summarize this decomposition in \autoref{tab:extension} and turn now to discuss each feature in more detail. Note that any calculations involving the \nee\ utilize the \widechan\ cube due to the higher SNR. However, the results do not change significantly if we instead use the \narchan\ cube.

\underline{\textsc{\nea}}: This component has a peak SNR $> 3$ in both the \narchan\ and \widechan\ data cubes. However, in the channel maps it is only marginally detected at $\sim 3\sigma_{\rm rms}$ over a small number of channels. Note that what appears to be a low-level feature at $\sim$ 199 km s$^{-1}$ in the \narchan\ cube is blended into \nea\ in the \widechan\ cube. Inspecting the extension-axis PV diagrams in \autoref{fig:pv} and \autoref{fig:extension}, we find tentative evidence for a faint `bridge' of emission spanning $\sim 100\arcsec$ (2.7 kpc) and covering the range $200 - 215$ km s$^{-1}$, of which the \nea~emission would represent the furthest extent. However, the sensitivity of our imaging precludes a definitive detection.

At lower velocities, there appears to be an \hi\ absorption feature, that in \narchan\ has a peak SNR of $\sim 2.7$. However, the channel maps at these velocities do not suggest any coherent structure. Furthermore, in the \widechan\ cube, which is more sensitive, the SNR remains below 3. There is also no obvious continuum or bright background \hi~in that region which would produce an absorption feature. We do not consider this feature any further.
 
 If the \nea\ is real, it has an \hi\ mass of $M_\subhi = (\mhineafive \pm \emhineafive) \times 10^5\rm\ M_\odot$.

\begin{deluxetable}{lccc}
\tablewidth{0pt}
\tablecaption{\nee\\ Decomposition}\label{tab:extension}
\tablehead{ 
Component	&	\colhead{$v_{\rm peak}^a$}	&	\colhead{$S_{\rm peak}^b$}	&	SNR \\
			&	(km s$^{-1}$)				&	(mJy)					& 	\\ \vspace{-0.4cm}  
}
\startdata
\multicolumn{4}{c}{\narchan\ rms: 1.8 mJy} \\
\hline
\nea			&	205						&	6.3						&	3.5 \\
\neb			&	233						&	7.2						&	4.0 \\
\nec			&	249						&	4.5						&	2.5 \\
\ned			&	276						&	4.7						&	2.6 \\
\hline
\multicolumn{4}{c}{\widechan\ rms: 0.7 mJy} \\
\hline
\nea			&	205						&	4.7						&	6.7 \\
\neb			&	233						&	5.2						&	7.4 \\
\nec			&	249						&	4.6						&	6.6 \\
\ned			&	277						&	5.2						&	7.4 \\
\enddata
\tablenotetext{a}{Channel with the brightest emission closest to the peak of the best fit Gaussian.}
\tablenotetext{b}{Peak flux density of the best fit Gaussian.}
\end{deluxetable}

\underline{\textsc{\neb}}: This component of the \nee\ recurs across multiple channels in all data cubes, and the shape of the emission in the various extension-axis PV diagrams (marked in blue in \autoref{fig:extension}) suggest that it is the most spatially connected emission in the region of the \nee\ (see also the central panels of \autoref{fig:chanmap}). From the \hi\ profile in \autoref{fig:extension}, we estimate an \hi\ mass of $M_\subhi = (\mhinebfive \pm \emhinebfive) \times 10^5\rm\ M_\odot$.

\underline{\textsc{\nec}}: This component is not extremely persistent, occurring over only a handful of channels at $2\sigma_{\rm rms}$ in \autoref{fig:chanmap}. While it is not obvious in the PV diagrams in \autoref{fig:extension}, it remains a confident detection in the more sensitive \widechan\ cube, as shown in \autoref{tab:extension}. Its inferred \hi\ mass is $M_\subhi = (\mhinecfive \pm \emhinecfive) \times 10^5\rm\ M_\odot$.

\underline{\textsc{\ned}}: This is the most kinematically distinct component of the \nee, separated by $\sim 40\rm\ km\ s^{-1}$ from the systemic velocity of Pisces A. It is detected at $>3\sigma_{\rm rms}$ in all data cubes, and is consistent across multiple channels in \autoref{fig:chanmap}, suggesting a coherent structure. The mass associated with this clump is $M_\subhi = (\mhinedfive \pm \emhinedfive) \times 10^5\rm\ M_\odot$.

\subsection{Beyond Bulk Motion: \\\ Marginal Detections}\label{sub:misc}

\begin{figure*}
\gridline{\hspace{-1cm}
\fig{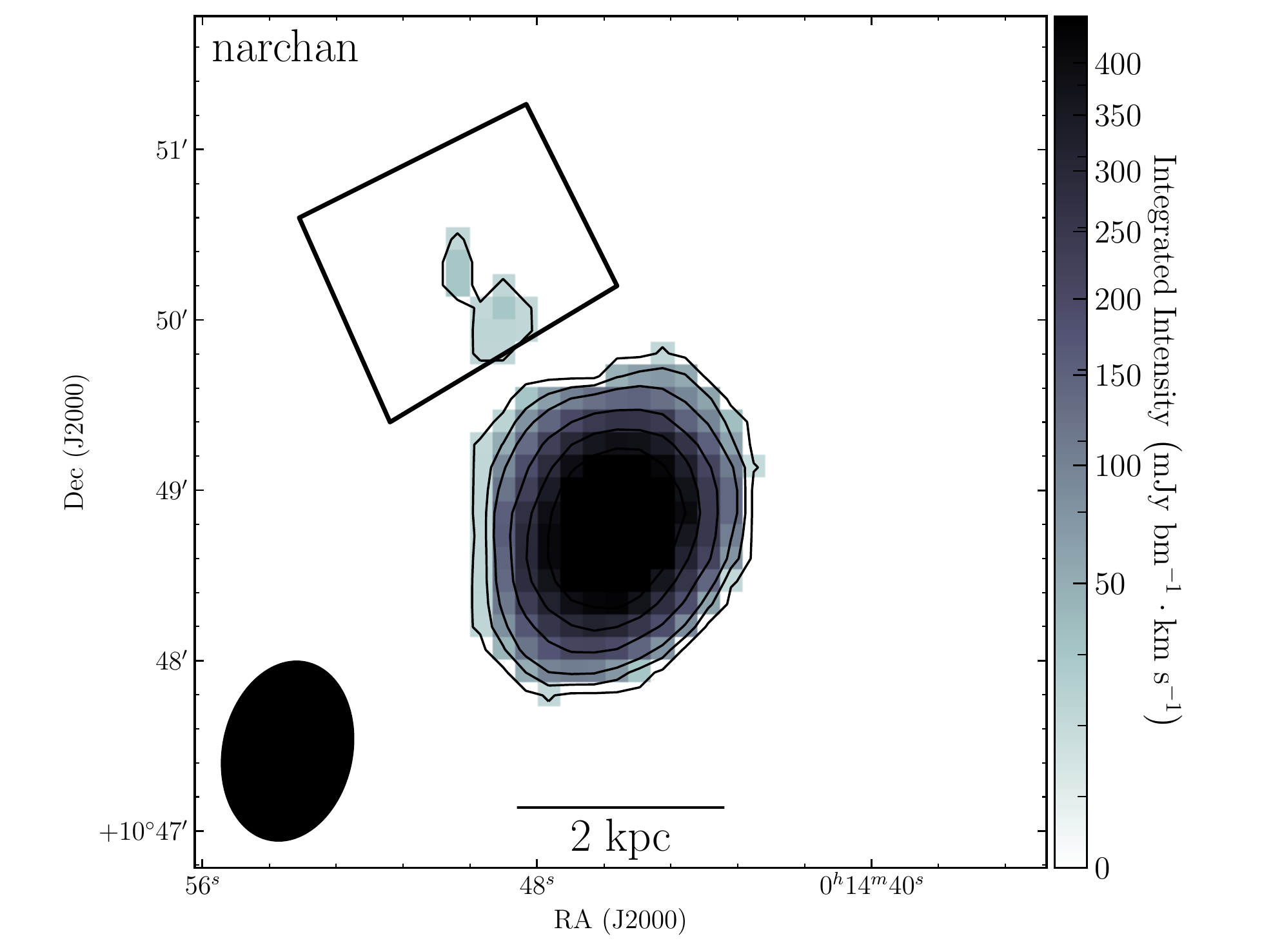}{0.37\linewidth}{}\hspace{-0.6cm}
\fig{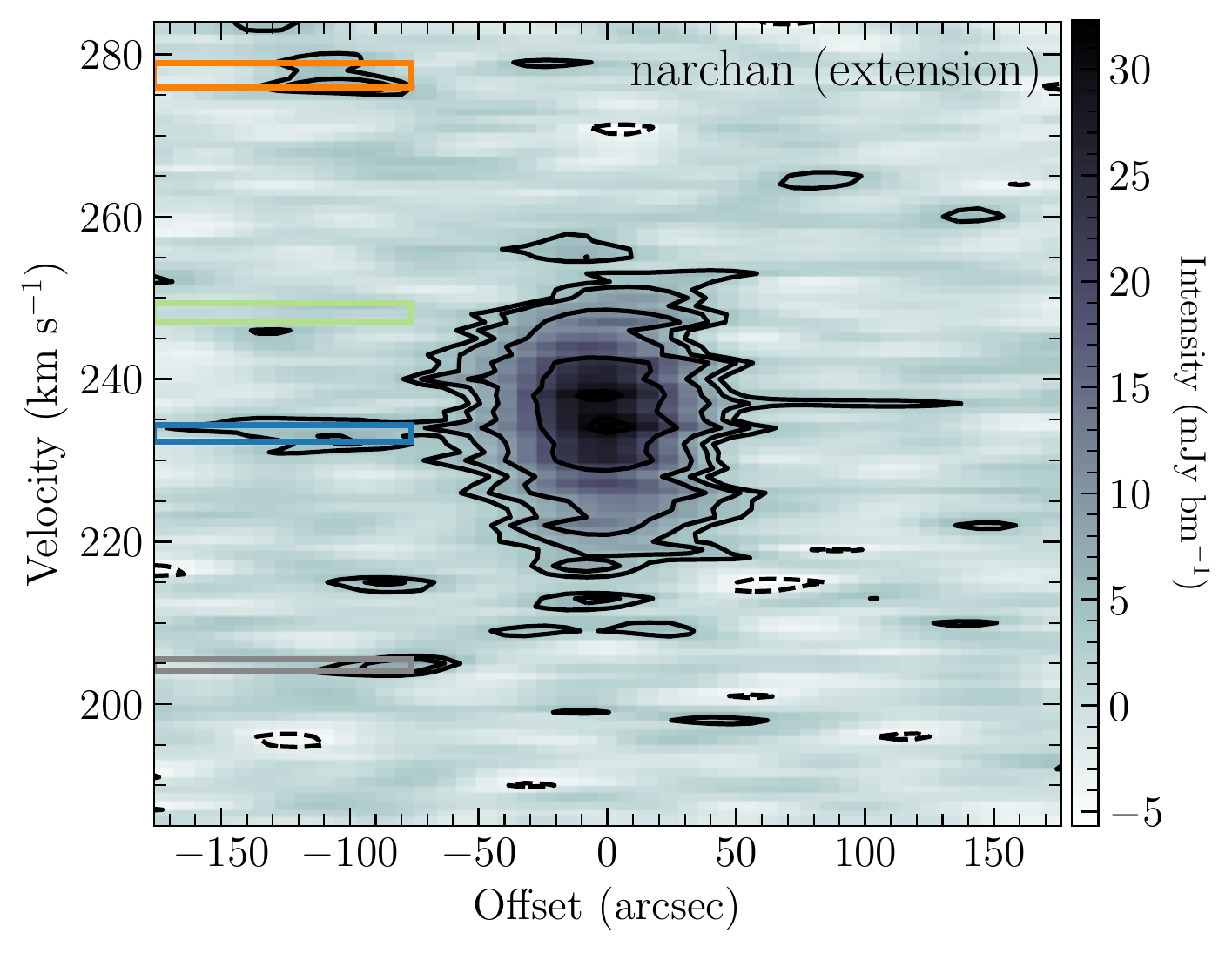}{0.36\linewidth}{}\hspace{-0.2cm}
\fig{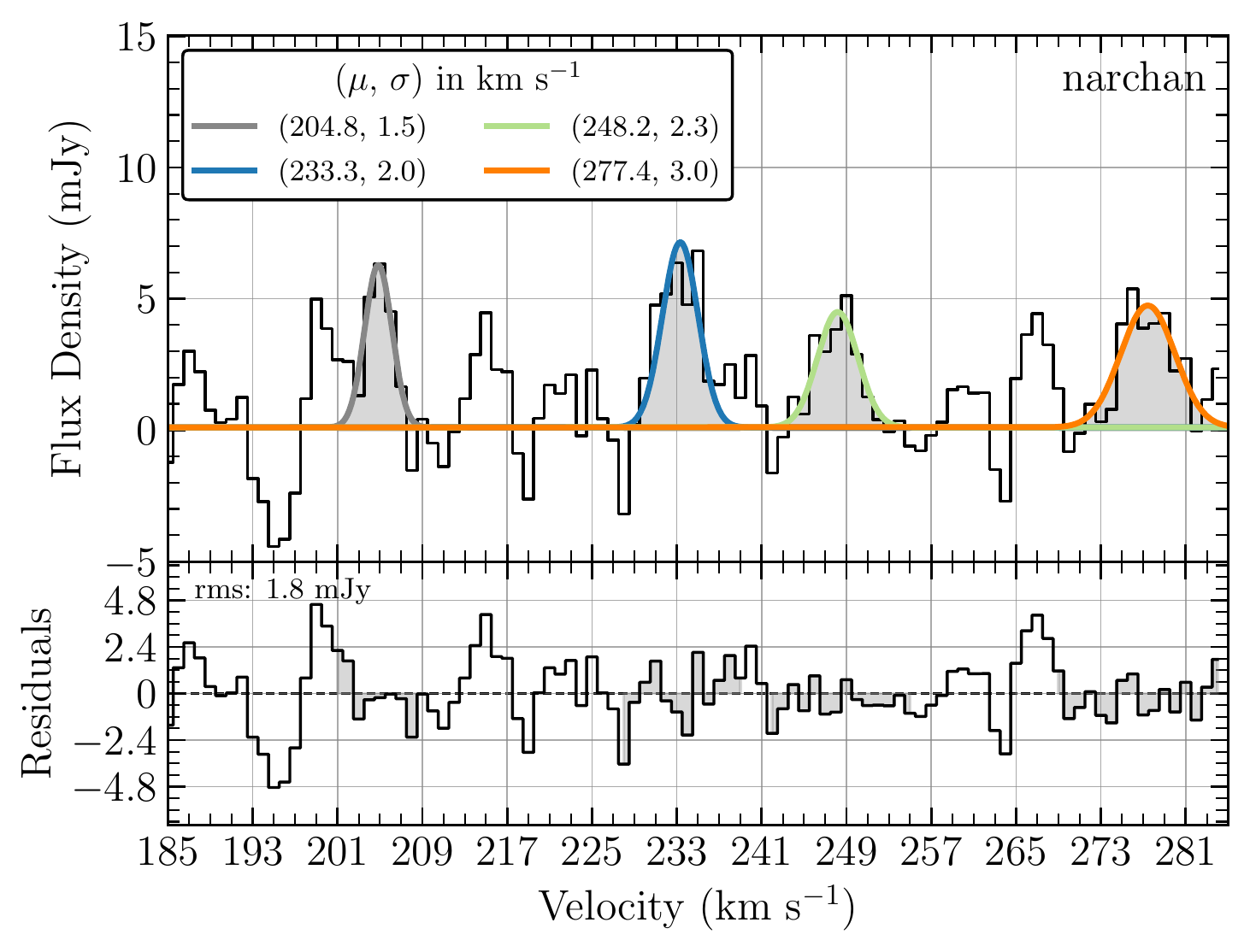}{0.37\linewidth}{}
}\vspace{-1cm}
\gridline{\hspace{-1cm}
\fig{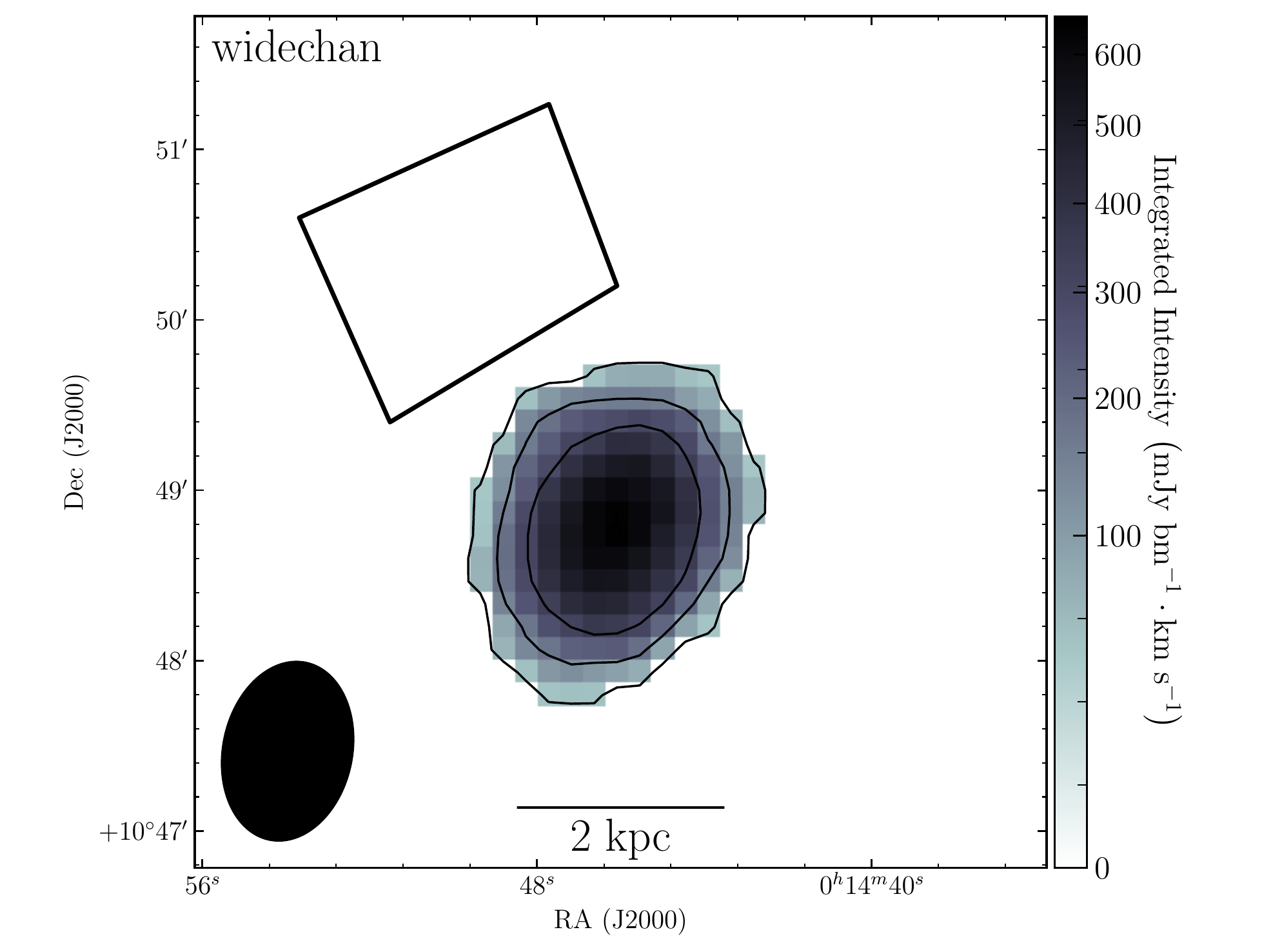}{0.37\linewidth}{}\hspace{-0.6cm}
\fig{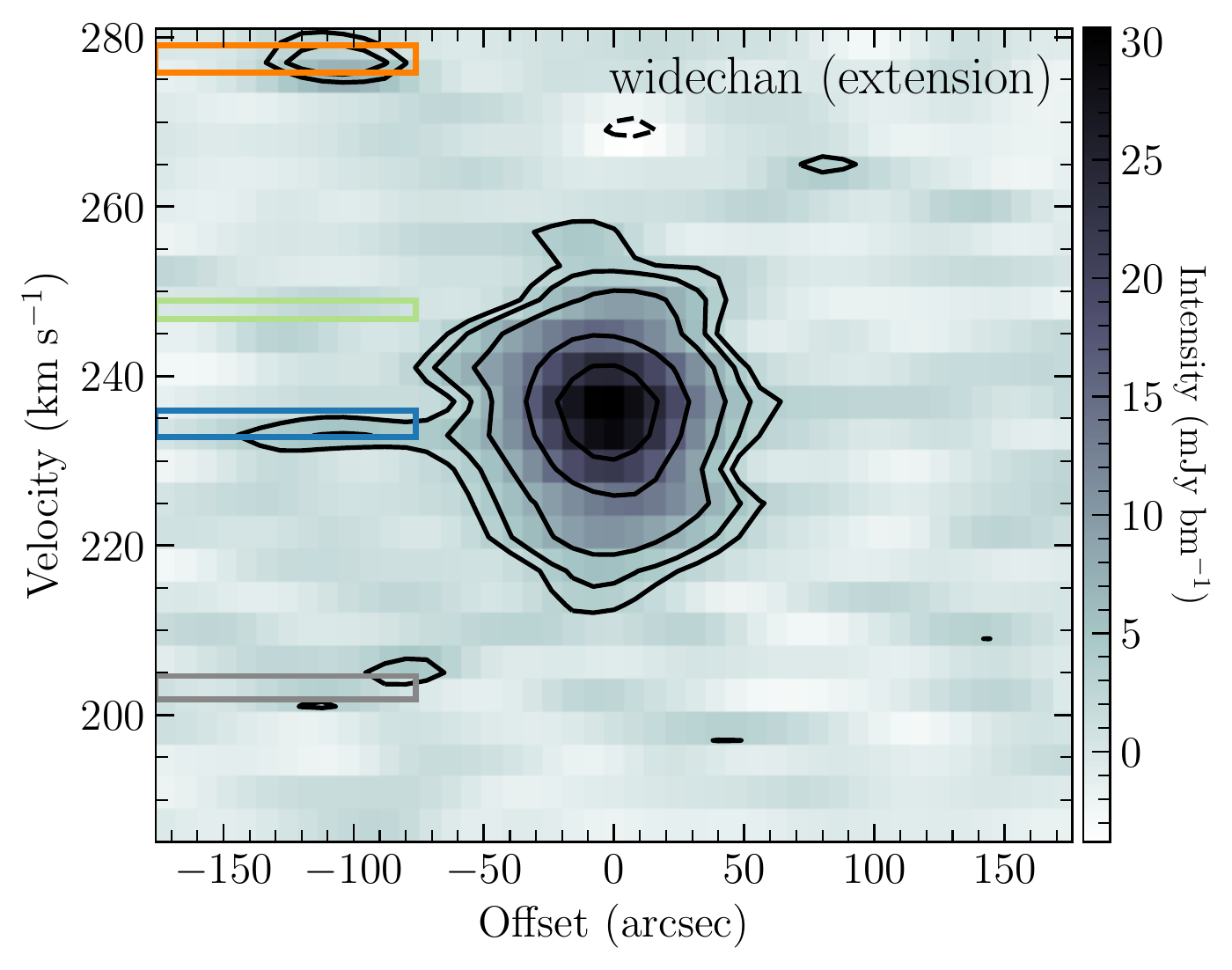}{0.36\linewidth}{}\hspace{-0.2cm}
\fig{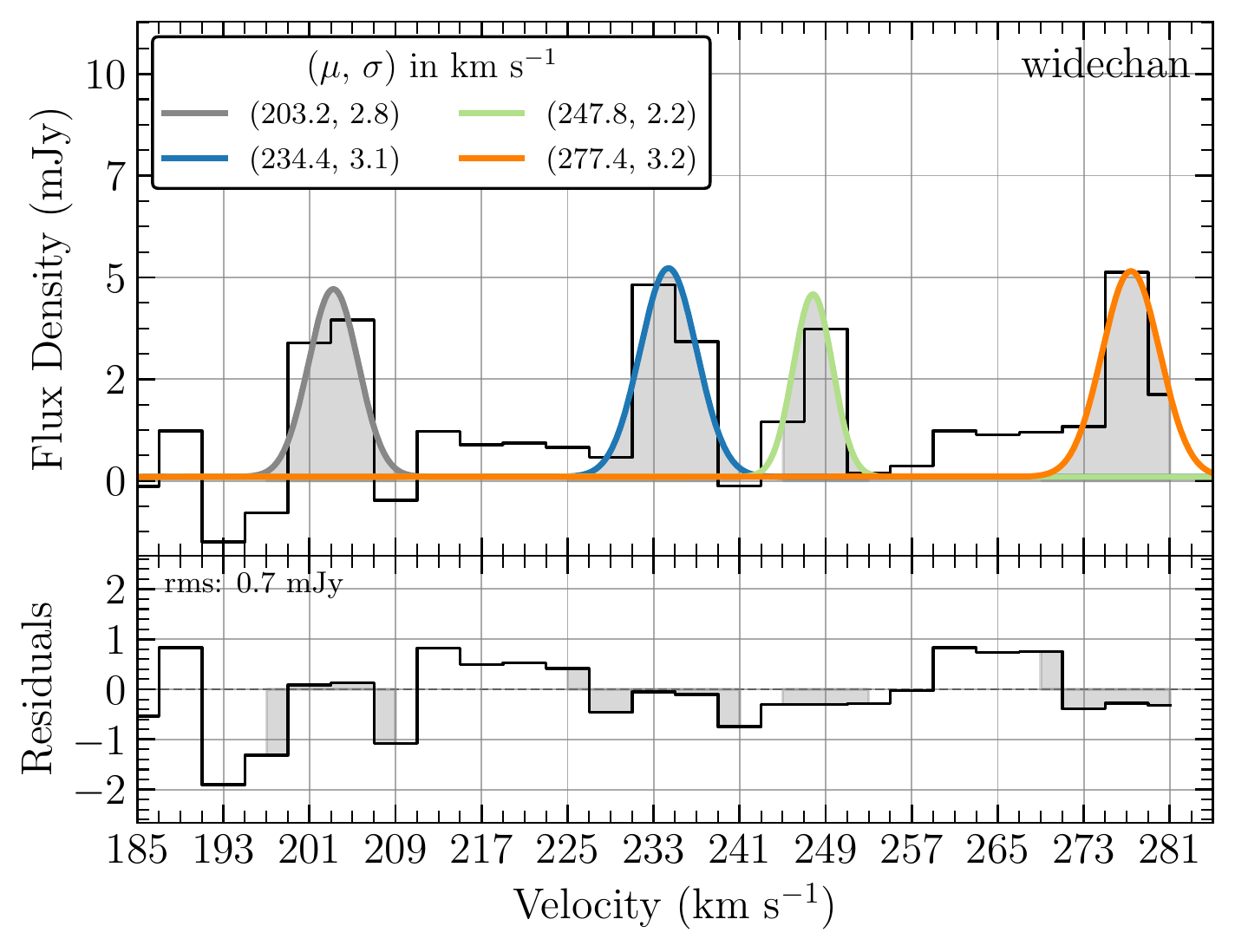}{0.37\linewidth}{}
}\vspace{-0.5cm}
\caption{\small Kinematic decomposition of the \nee\ in the \narchan\ (top) and \widechan\ (bottom) data cubes. \textit{Left}: Moment 0 map showing the most compact component of the extension. The black trapezium outlines the region used to extract the global spectrum. Contours are at (1, 5, 10, 20, 30, 40) $\times 10.2 (\times 32)$ mJy bm$^{-1} \cdot\rm km~s^{-1}$ for the \narchan\ (\widechan) cube. \textit{Center}: Extension-axis PV diagram with contours similar to \autoref{fig:pv}. The colored rectangles bound the spatial and spectral extent of the best fit Gaussians. \textit{Right}: Corresponding global spectrum of the \nee. Thick colored lines represent Gaussian fits to the brightest components. Colors are the same as the rectangles in the center panels.}
\label{fig:extension}
\end{figure*}

The channel maps presented in \autoref{fig:chanmap} reveal a possible feature that we call the ``southwest extension'', and it is most obvious at 237 km s$^{-1}$. In the PV diagrams of \autoref{fig:pv} and \autoref{fig:extension} (particularly in the \narchan\ cube), this is seen as a very narrow feature weakly connected to the main body of the galaxy.  It is important to note that while we see it at similar velocities to the \nee, particularly \neb, it is not as persistent. Additionally, it is nearly spatially coincident with continuum sources and background galaxies (see \autoref{fig:hst+cont} and \autoref{fig:moment}). We return to this feature in our discussion of possible tidal forces below.

In \autoref{tab:clumps}, we summarize the \hi\ properties of all anomalous \hi\ features. Separations are calculated with respect to the optical center of Pisces A, and both separations and sizes assume the clump is at the same distance as the galaxy itself.

\subsection{Dynamics \& Baryon Fraction}\label{sub:dynamics}

\begin{figure}
\gridline{\vspace{-0.8cm}
\fig{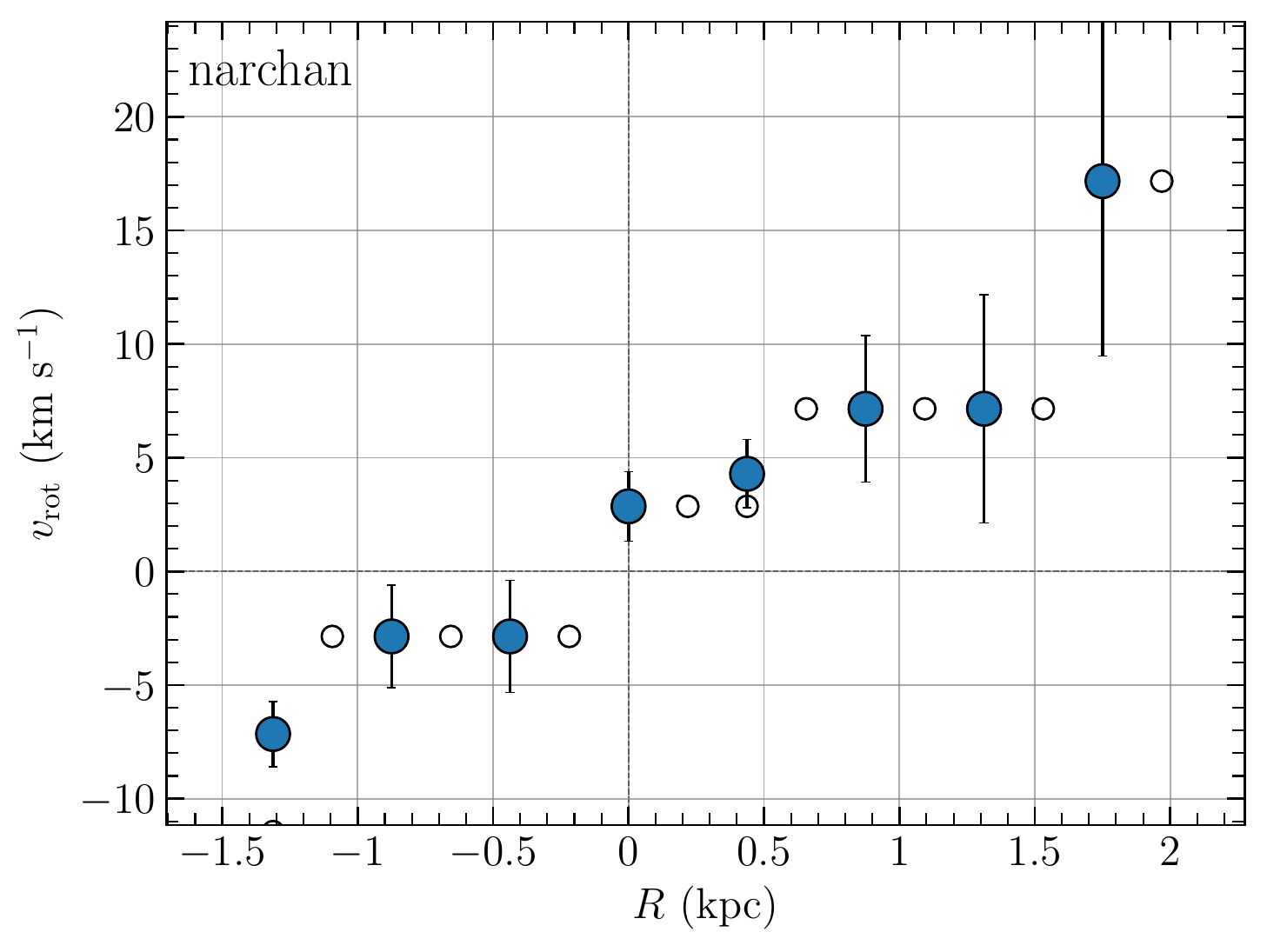}{\linewidth}{}
}
\caption{\small Unfolded inclination-corrected rotation curve of Pisces A from the \narchan\ data cube. Open circles denote the $>2\sigma_{\rm rms}$ oversampled pixels. The final resampled rotation curve is marked by the blue filled circles. Negative galactocentric distances indicate the approaching side. See Section \ref{sub:rotcur} for more details on the sampling and error estimation.}
\label{fig:rotcur}
\end{figure}

\begin{figure}
\gridline{\vspace{-0.8cm}
\fig{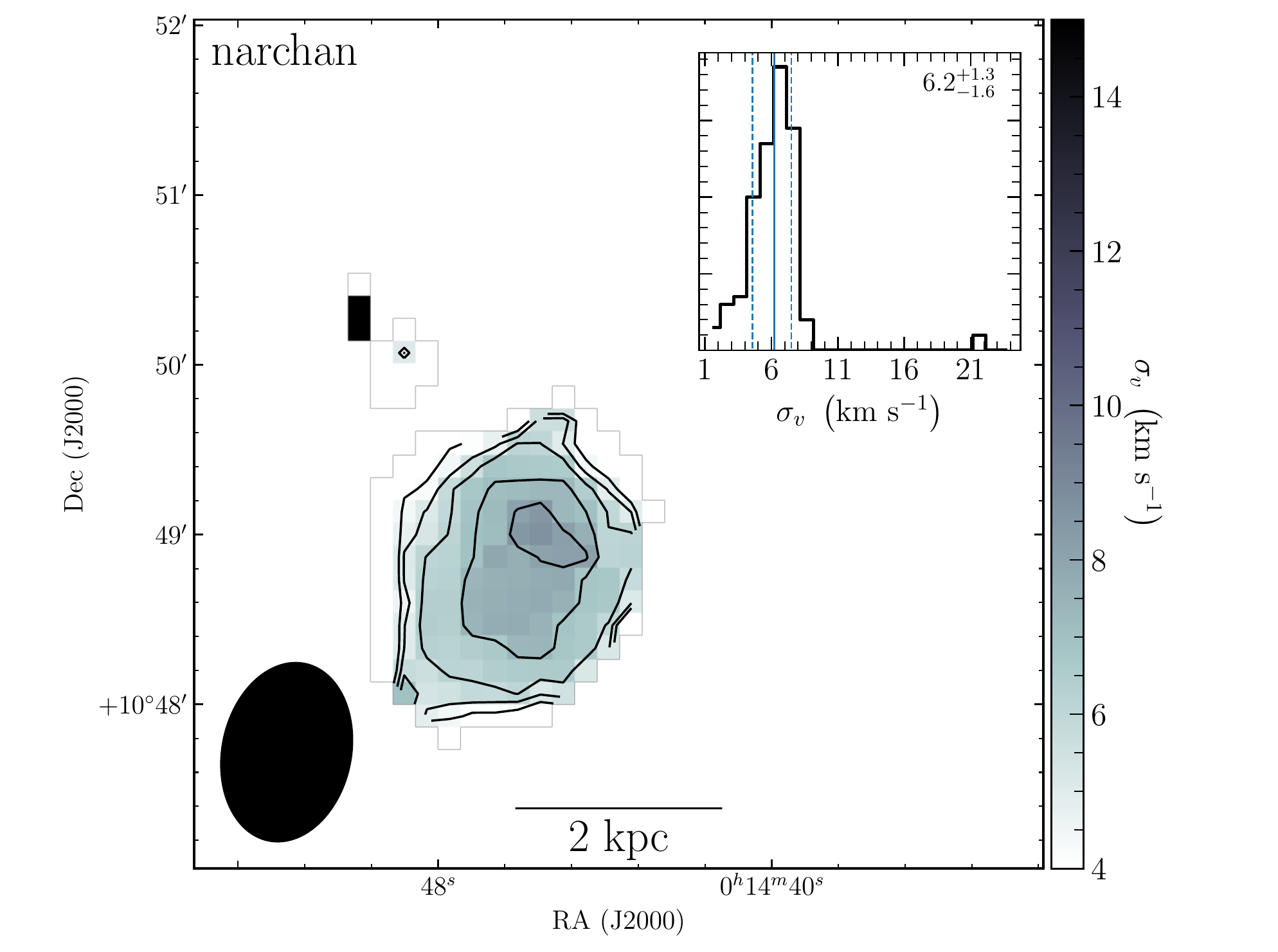}{\linewidth}{}
}
\caption{\small Dispersion map from the \narchan\ data cube. Contours are spaced at 1 km s$^{-1}$ intervals. \textit{Inset}: Corresponding distribution of velocity dispersions.}
\label{fig:dispersion}
\end{figure}

In \autoref{fig:rotcur} we present the rotation curve of Pisces A, derived from the \narchan\ data cube in order to best characterize the rotation. It is clear that the rotation curve is still rising out to the last measured data point, indicating that we are only detecting the solid body rotation of the galaxy. This is expected considering the beam size relative to Pisces A. Rotation curve fitting, which would provide insight into the dark matter profile, is beyond the scope of this paper. However, we may still compute the baryon fraction in Pisces A, and thus estimate the dark matter content.

Because we do not have a complete knowledge of the matter distribution in the galaxy, i.e., the detailed interplay between rotation and dispersion, we do not have a direct method of calculating the total dynamical mass. Instead, we use the following approximation:

\begin{equation}\label{eqn:dynamical}
M_{\rm dyn} = 2.325 \times 10^5\ {\rm M_\odot} \left( \frac{v_{\rm rot}^2 + 3\sigma^2}{\rm km^2\ s^{-2}} \right) \left( \frac{R}{\rm kpc} \right),
\end{equation}

where $v_{\rm rot}$ is the rotational velocity, $\sigma$ is the velocity dispersion, and $R$ is the galactocentric distance. This functional form is a compromise between an isothermal sphere (with isotropic velocity dispersion) and a uniform density sphere (with isotropic and constant dispersion), as determined by applying the virial theorem to dwarf galaxies \citep[see][and references therein]{hoffman+1996}.

\autoref{fig:dispersion} shows the dispersion map and corresponding distribution of values, also derived from \narchan. We find a velocity dispersion of $\sigma = \vdisp^{+ \evdisphi}_{- \evdisplo}$ km s$^{-1}$, where the errors represent the inner 68\% of the values.

Assuming $v_{\rm rot} = \vrot \pm \evrot$ km s$^{-1}$ and $R = \rrot$ kpc from the approaching side of \autoref{fig:rotcur}, we find a total dynamical mass of $M_{\rm dyn} = (\mdyneight \pm \emdyneight) \times 10^8\rm\ M_\odot$. It is important to stress that this is a lower limit, since we have not captured the full rotation curve. Additionally, the effects of smoothing act to flatten the rotation curve, further underestimating $v_{\rm rot}$. Assuming the total baryonic mass is $M_{\rm bary} = M_\star + 1.4 M_{\subhi,\rm tot}$, we find a baryon fraction of

\begin{equation}\label{eqn:fbary}
f_{\rm bary} = \frac{M_{\rm bary}}{M_{\rm dyn}} = \totalbaryonfraction \pm \ebaryonfraction
\end{equation}

Excluding all the components of the northeast extension from the baryonic mass budget does not change $f_{\rm bary}$ in any meaningful way.

\section{Pisces A in Context}\label{sec:context}

\subsection{The Cosmic Environment}\label{sub:environment}

\citet{tollerud+2016} highlighted the fact that Pisces A appears to lie near the boundary of local filamentary structure (see their Section 6). Using a sample of galaxies within the Local Volume synthesized from the NASA-Sloan Atlas \citep{blanton+2011}, the Extragalactic Database \citep{tully+2009}, the 6dF Galaxy Survey \citep{jones+2009}, and the 2MASS Extended Source Catalog \citep{skrutskie+2006} and volume-limited to the detection limit of 2MASS at 10 Mpc, $M_K < -17.3$, they showed that Pisces A is located in an overdensity which appears to point toward the Local Void \citep[18$^h$38$^m$ 18$^\circ$;][]{karachentsev+2002}.

We investigate this idea further by characterizing the underlying density field. We employ the basic method of \citet{sousbie2011}, which takes advantage of the fact that the structures of the cosmic web (voids, walls, filaments) have well-defined analogues in computational topology. While a full description of the method is beyond the scope of this paper, we briefly summarize it here.

Given a discrete sampling of a topological space (e.g., a sample of galaxies), we may compute the Delaunay tessellation, which is a triangulation of the points that maximizes the minimum interior angle of the resulting triangles \citep{delaunay1934}. Each point is a vertex of $N$ Delaunay triangles (each with area $A_{\rm Delaunay}$), and we may treat the sum of those triangles as the ``contiguous Voronoi cell'', with area $A_{\rm Voronoi} = \sum_{i=1}^N A_{{\rm Delaunay}, i}$. Intuitively, the larger the contiguous Voronoi cell, the more isolated the point. Quantitatively, we say the area $A_{\rm Voronoi}$ is inversely related to the density, and so we assign an estimate of the density to each point as $\rho \propto A_{\rm Voronoi}^{-1}$. Finally, we reconstruct the full (continuous) density field by interpolating between estimates.

\begin{figure}
\includegraphics[width=\linewidth]{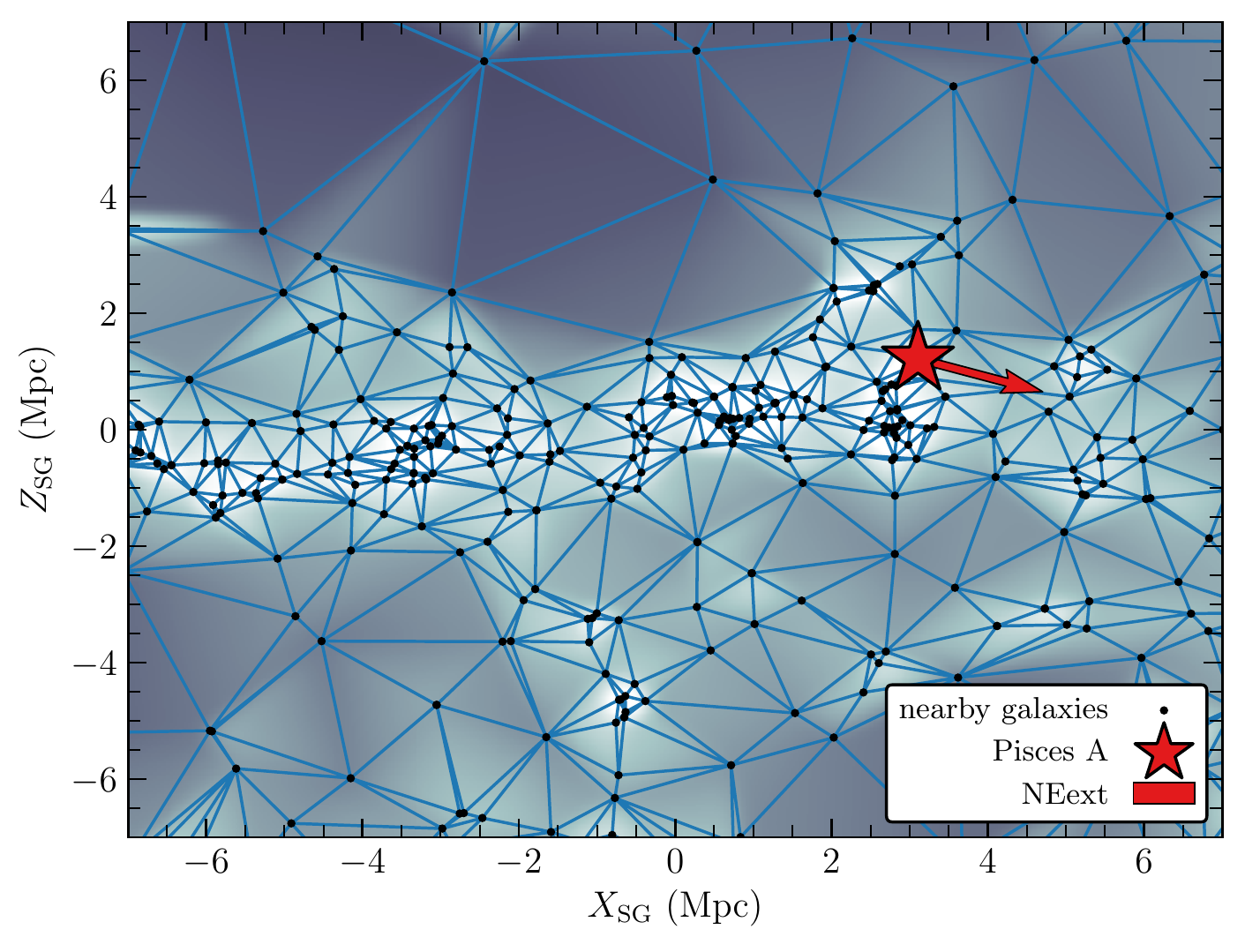}
\caption{\small The cosmic density field around Pisces A. Black points are a sample of nearby galaxies \citep[following the prescription in][]{tollerud+2016}. Pisces A is shown as a red star, with the \nee~pointed in the direction of the red arrow (not to scale). The Delaunay tessellation is given in blue, with the corresponding density field in greyscale.}
\label{fig:localvoid}
\end{figure}

The Delaunay tessellation of galaxies in the Local Void (and the estimated underlying density field) is presented in \autoref{fig:localvoid} in supergalactic Cartesian coordinates. In this coordinate system, the viewer is at the origin. Black points denote nearby galaxies (as described above). Pisces A is overplotted as a red star, with an arrow (not to scale) marking the projected direction of the \nee~onto the $X_{\rm SG}-Z_{\rm SG}$ plane. As this is a simplified version of the method of \citet{sousbie2011}, our reconstructed cosmic web has only been minimally smoothed for visualization, and thus artificial boundaries still exist. However, it is immediately clear that there is filamentary structure in the underlying density field, beyond the appearance of the data themselves. Pisces A is well within this filament, supporting the idea of its current membership. In this projection, the \nee~appears nearly parallel to the filament, hinting at a possible relationship.

We consider the full 3D distribution in \autoref{fig:localvoid2}. This filament has been previously recognized by \citet{tully+2008}, which they labeled the ``Local Sheet'' due to its pancake-like structure. Its cylindrical boundaries (7 Mpc in radius and 3 Mpc thick) are plotted as thick black lines. Nearby galaxies are plotted as circles colored by their peculiar velocity with respect to the Local Sheet, defined as $v_{\rm pec} = v_{\rm LS} - H_0 d$, where $v_{\rm LS}$ is the heliocentric velocity transformed to the Local Sheet frame (their Equations 15 and 16), $d$ is the distance, and $H_0 = 74.03$ km s$^{-1}$ Mpc$^{-1}$ \citep{riess+2019}. Pisces A has a peculiar velocity of $v_{\rm pec} \approx 19$ km s$^{-1}$, compared with the median peculiar velocity of $21$ km s$^{-1}$ for galaxies in our sample that are within 2 Mpc of the boundary of the Local Sheet. In this 3D view, the gas extension appears to point mostly within the plane of the Local Sheet, nearly tangential to the radial boundary. We remind the reader that the arrows denote the projected direction of the \nee~but not its motion or length, which at this scale would be smaller than the marker for the galaxy.

\subsection{The Dwarf Galaxy Population}\label{sub:populations}

\begin{deluxetable*}{lcccc}
\tablecaption{Derived \hi~Clump Properties}\label{tab:clumps}
\tablehead{
								&	\multicolumn{4}{c}{\nee~(\widechan)} \\
									\cmidrule(lr){2-5}  
Parameter							&	\colhead{\nea}		&	\colhead{\neb}		&	\colhead{\nec}		&	\colhead{\ned}
}
\startdata
$v_{\rm fit}$ (km s$^{-1}$)				&	$203.2 \pm 1.4$	&	$234.4 \pm 1.4$	&	$247.8 \pm 5.0$	&	$277.4 \pm 1.5$ \\
$v_{\rm peak}$ (km s$^{-1}$)$^a$		&	205				&	233				&	249				&	277 \\
Projected Separation (kpc)			&	2.2				&	2.9				&	3.0				&	3.1 \\
Projected Size (kpc$\times$kpc)$^b$	&	$1.3 \times 0.9$	&	$1.6 \times 0.7$	&	$0.9 \times 0.6$	&	$1.9 \times 1.2$ \\
$\Delta v_{\rm fit}$ (km s$^{-1}$)$^c$	&	$-32.5 \pm 1.6$		&	$-1.3 \pm 1.6$		&	$12.1 \pm 5.0$		&	$41.7 \pm 1.5$ \\
$\Delta v_{\rm peak}$ (km s$^{-1}$)$^c$	&	$-30.7$			&	$-2.7$			&	$13.3$			&	$41.3$ \\
$W50^i_\subhi$ (km s$^{-1}$)			&	$9.4 \pm 14.2$		&	$10.5 \pm 7.5$		&	$7.4 \pm 21.9$		&	$10.8 \pm 5.2$ \\
$M_\subhi$ (10$^5$ M$_\odot$)		&	$2.1 \pm 1.4$		&	$2.6 \pm 1.6$		&	$1.5 \pm 1.2$		&	$2.4 \pm 1.4$ \\
\enddata
\tablenotetext{}{\hspace{3.3cm}$^a$Channel with the brightest emission. \\ \hspace*{3.45cm}Also used to measure the separation and size of the clump.}
\tablenotetext{}{\hspace{3.3cm}$^b$Based on the $2\sigma_{\rm rms}$ contour.}
\tablenotetext{}{\hspace{3.3cm}$^c$Defined as $v - v_{\rm sys, \subhi}$ (see Table \ref{tab:properties}).}
\end{deluxetable*}

\begin{figure*}
\includegraphics[width=\linewidth]{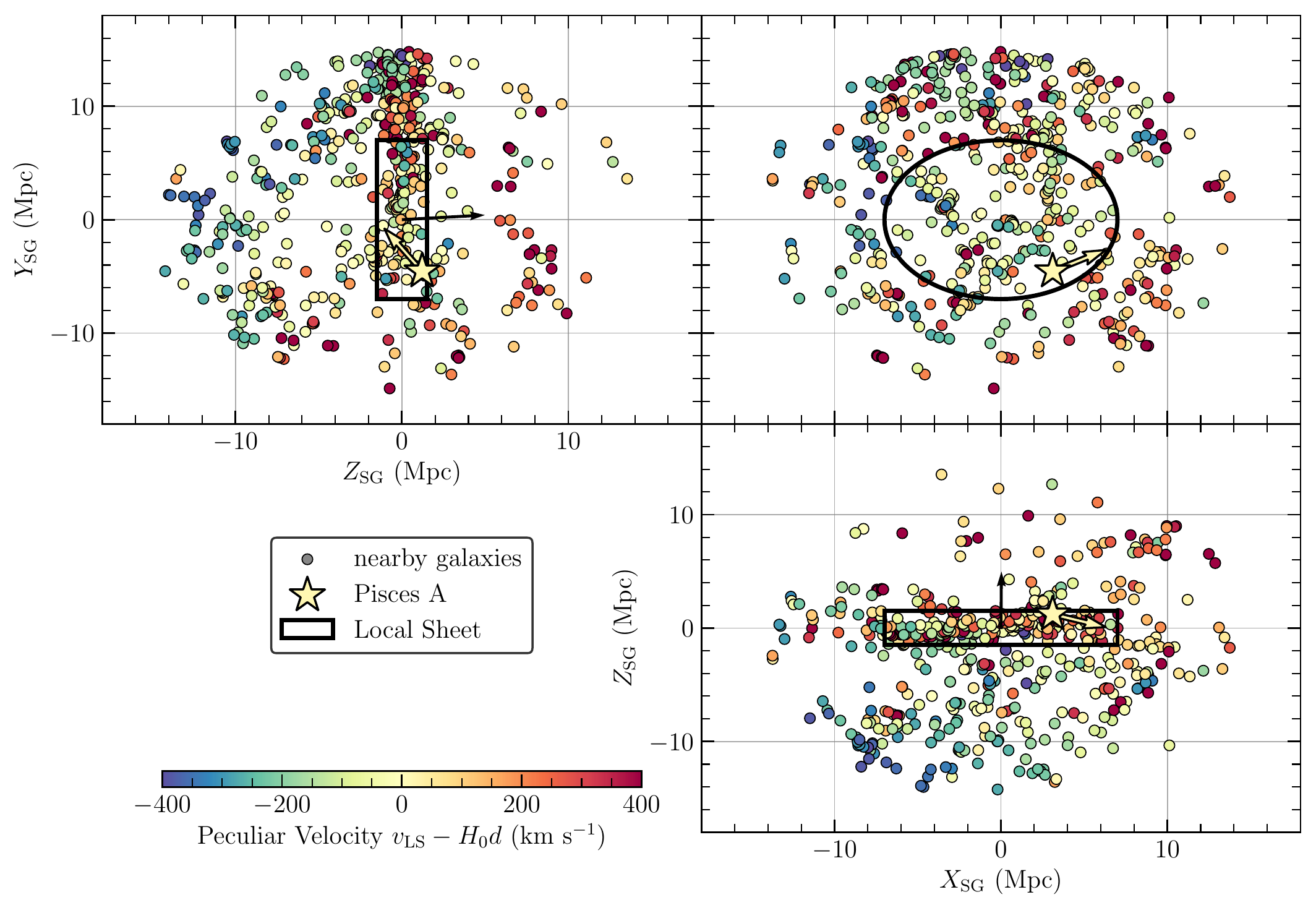}
\caption{\small Galaxies in the Local Volume. Data points and coordinate system are the same as \autoref{fig:localvoid}, but now for all three coordinate planes. The black arrow points toward the Local Void. In the $X_{\rm SG}-Y_{\rm SG}$ plane (upper right panel), this is into the page. The yellow arrow (not to scale in any panel) denotes the projected direction of the \nee~and does not represent its 3D motion, nor the motion of the galaxy. Points are colored according to their peculiar velocity with respect to the Local Sheet \citep{tully+2008}, assuming $H_0 = 74.03$ km s$^{-1}$ Mpc$^{-1}$ \citep{riess+2019}. The thick black lines represent the boundary of the Local Sheet.}
\label{fig:localvoid2}
\end{figure*}

\begin{figure}
\includegraphics[width=\linewidth]{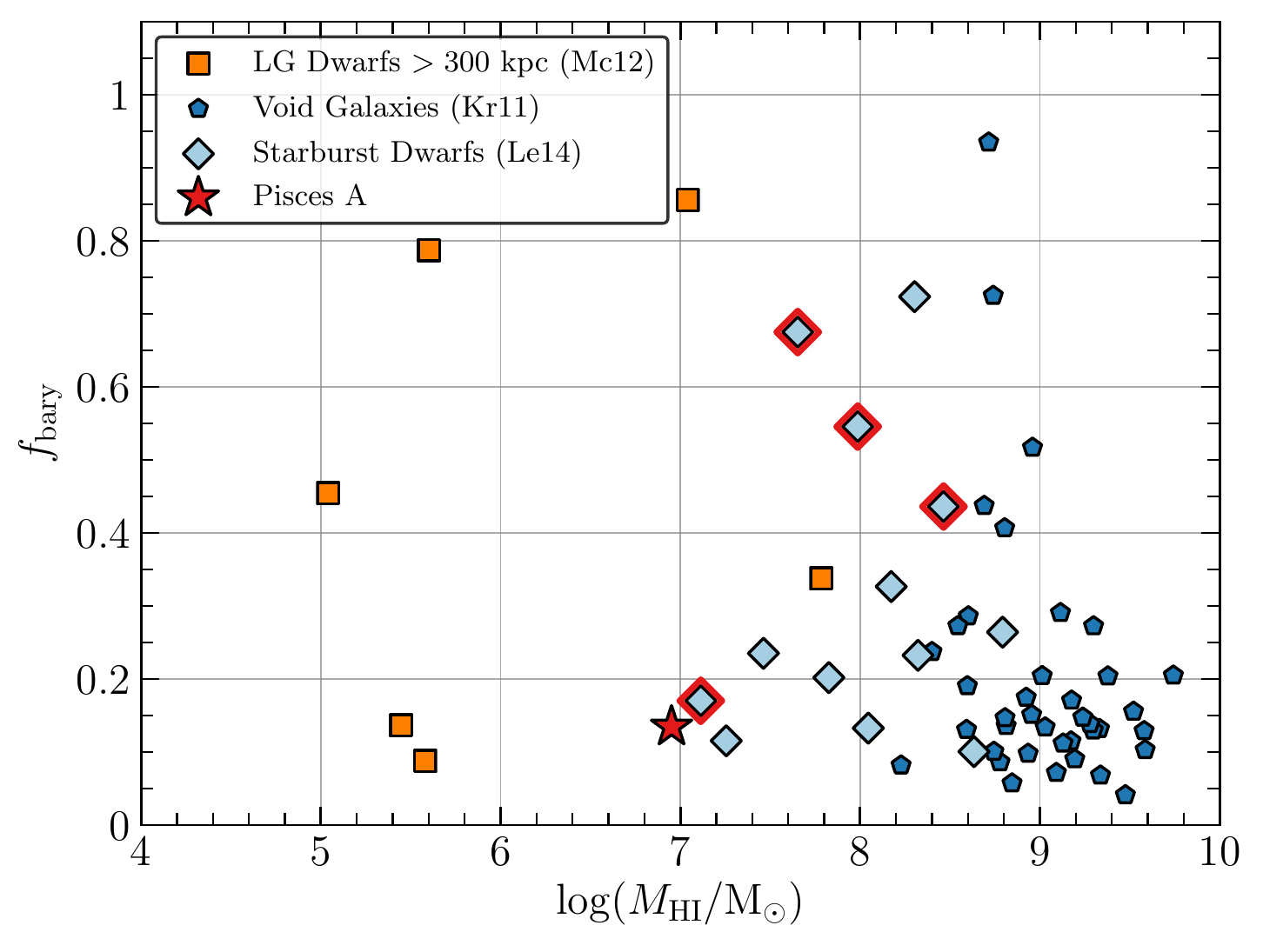}
\caption{\small The $\log M_\subhi-f_{\rm bary}$ plane. Pisces A is plotted as a red star. Orange squares are a sample of dwarf galaxies in the Local Group outside of the virial radius of the Milky Way that have dynamical masses. Blue pentagons are a sample of void galaxies from the Void Galaxy Survey \citep{kreckel+2011a}. Light blue diamonds are local starburst dwarf galaxies \citep{lelli+2014b}. Those outlined in red are considered to have kinematically disturbed \hi\ disks as defined in \citet{lelli+2014a}.}

\label{fig:fbarmhi}
\end{figure}

In \autoref{fig:fbarmhi}, we place Pisces A in the $\log M_\subhi-f_{\rm bary}$ plane. Also plotted are a sample of Local Group (LG) galaxies from \citet{mcconnachie2012} that have measured rotational velocities and are selected to exist beyond the virial radius of the Milky Way. Where available, the dynamical mass is reported as $M_{\rm dyn} \propto r_h \sigma_\star^2$ where $r_h$ is the half-light radius and $\sigma_\star$ is the stellar velocity dispersion. Only 6 galaxies (NGC205, NGC185, LGS3, Leo T, WLM, and Leo A) have an \hi\ detection and available dynamical information. We also include galaxies from the Void Galaxy Survey \citep{kreckel+2011a, kreckel+2014}, which are selected to reside in the deepest underdensities within the SDSS. Here, $M_{\rm dyn}$ is estimated from $W50_\subhi$ (where available) and $R_\subhi$, the radius at which the surface density $< 1\rm\ M_\odot\ pc^{-2}$. Finally, we include a sample of nearby starburst dwarf galaxies selected to have resolved \textit{HST} observations \citep{lelli+2014b}. Note that those galaxies marked as having kinematically disturbed \hi\ disks do not have reliable rotation curves -- instead, the kinematical parameters are estimated from the outermost parts. We estimate the dynamical mass following \autoref{eqn:dynamical}, assuming a velocity dispersion $\sigma = 10\rm\ km\ s^{-1}$ (see their Table 4). Pisces A has a baryon fraction that is not inconsistent with the Local Group dwarfs or starbursts but is most similar to the bulk of the void galaxy population, just at a lower mass.

\citet{kreckel+2011a} find that void galaxies tend to be gas-rich with ongoing accretion and alignment with filaments. Although Pisces A is currently embedded within the Local Sheet (see \autoref{fig:localvoid2}), it is $\lesssim 300$ kpc from the top edge of the structure. We do not know the total space motion of the galaxy. However, its proximity to the edge of the Local Void, coupled with the aforementioned similarity to known void galaxies and relatively recent change in star formation rate, suggests that Pisces A may have originated within the Local Void and has since transitioned into, and achieved equilibrium relative to, a denser environment.

Such a scenario is reminiscent of KK 246, a low-mass dwarf ($M_\subhi \sim 10^8\rm~M_\odot$) confirmed to be located within the Local Void \citep{kreckel+2011b}. \citet{rizzi+2017} use a numerical action model combined with an updated distance to show that this galaxy is moving rapidly away from the void center (toward us) as a result of the expansion of the void. It is possible that Pisces A has made this journey already, which is supported by its peculiar velocity matching very well the overall motion of the Local Sheet.

\begin{figure}
\includegraphics[width=\linewidth]{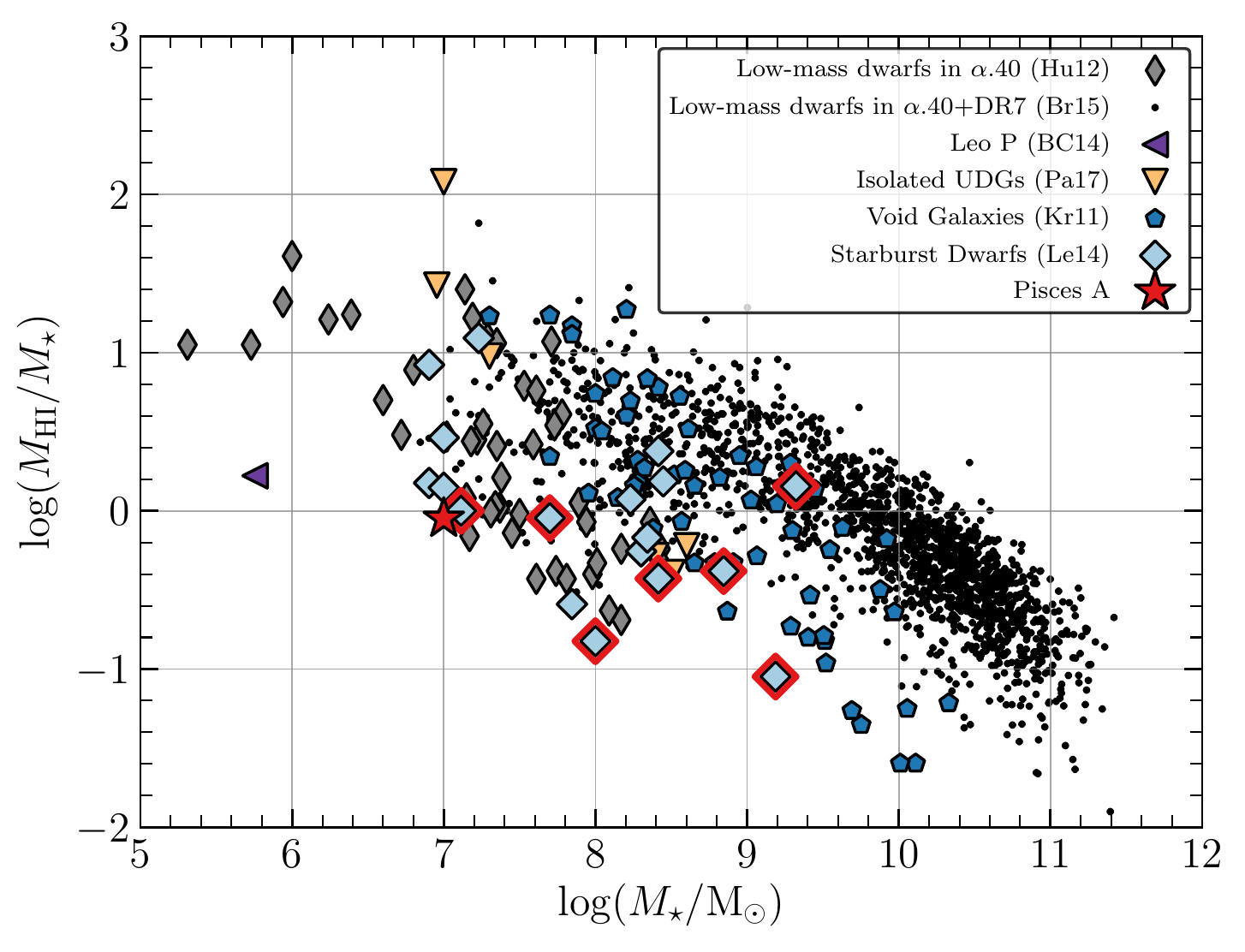}
\caption{\small The $\log M_\star-\log\left( M_\subhi / M_\star \right)$ plane. Pisces A is plotted as a red star. Thin grey diamonds represent low-mass dwarfs already discovered within the ALFALFA $\alpha$.40 catalog \citep{huang+2012}. Small black dots are from a comparison of low-mass isolated galaxies in SDSS to galaxies in $\alpha$.40 \citep{bradford+2015}. Leo P is displayed as a purple left-facing triangle \citep{bernstein-cooper+2014}. Yellow downward-facing triangles are isolated ultra-diffuse galaxies \citep{papastergis+2017}. Blue pentagons and light blue diamonds are as in \autoref{fig:fbarmhi}.}

\label{fig:massfrac}
\end{figure}

\autoref{fig:massfrac} presents the \hi\ mass fraction as a function of the stellar mass. For context, low-mass dwarfs in a variety of environments are included. Pisces A is similar to other low-mass dwarfs observed by ALFALFA, as well as the nearby starbursting dwarf galaxies. We find that Pisces A is somewhat gas-rich\footnote{We define ``gas-rich'' to be $M_{\rm gas} > M_\star$ \citep[as in, e.g.,][]{mcgaugh2012}, which corresponds to $\log\left( M_\subhi / M_\star \right) > -\log(1.4) \approx -0.146$. \citet{bradford+2015} instead defines a ``gas-depleted'' threshold (see their Figure 5). By this measure, Pisces A would not be considered gas-depleted.} with an \hi\ fraction of $\log\left( M_\subhi / M_\star \right) \approx -0.076$. Excluding the \hi\ clumps yields $\log\left( M_\subhi / M_\star \right) \approx -0.122$. However, relative to other dwarfs of a similar stellar mass, Pisces A is among the least gas-rich, with an \hi\ fraction up to 2 orders of magnitude lower than comparison galaxies. In particular, while its \hi~fraction is consistent with void galaxies (blue pentagons in the figure), it is on the extreme low end of the scatter for void galaxies of a similar stellar mass.

\section{Discussion}\label{sec:discussion}

The variety of spatially and kinematically distinct components in the \hi\ distribution of Pisces A suggests that it is not in equilibrium. In particular, our results in \autoref{tab:clumps} suggest that there exist relatively large clumps (up to $50\%$ of the linear size of Pisces A itself) harboring almost 10\% of the gas mass as in the main body of the galaxy. We consider a variety of scenarios to explain the origin of these clumps, as well as the timing of the uptick in the SFH of Pisces A.

\subsection{Stripping from Ram Pressure}\label{sub:rampressure}

The shape of the \nee\ is somewhat ``tail''-like, suggestive of the typical morphology of gas that has been stripped by ram pressure \citep[e.g.,][]{mcconnachie+2007}. This mechanism is crucial in shaping the satellites of the Milky Way \citep{mayer+2006, grcevich+2009}, and galaxies within groups can be stripped by the intragroup medium \citep[e.g.,][]{bureau+2002, fossati+2019}. \citet{benitezllambay+2013} find that in the Local Group, ram pressure stripping due to the cosmic web is efficient at removing gas, especially for low-mass dwarfs due to their weaker potential. Given that Pisces A appears to be embedded within the cosmic web (as can be seen in \autoref{fig:localvoid} and \autoref{fig:localvoid2}), it may have experienced similar stripping, leading to the disturbed morphology we observe. However, because the galaxy does not appear to be part of a group (see below), we instead consider the intergalactic medium (IGM) as a potential source, which simulations have shown can not only strip the gas, but also kickstart star formation in otherwise quiescent galaxies \citep{wright+2019}.

If the IGM is indeed the stripping medium, then its density must satisfy $n_{\rm IGM} \gtrsim 2 \pi G \Sigma_{\rm tot} \Sigma_\subhi / \mu v_{\rm rel}^2$, where $\Sigma_{\rm tot}$ is the total (including stars) surface density of the galaxy, $\Sigma_\subhi$ is the column density of the \hi\, and $\mu = 0.75 m_p$ for fully ionized media \citep{gunn+1972}. A reasonable estimate of the total column density is $\Sigma_{\rm tot} = \Sigma_\subhi (1 + M_\star / M_\subhi) \approx 2 \Sigma_\subhi$ \citep[for an example, see][]{mcconnachie+2007}. In the \foundation\ cube, the $3\sigma_{\rm rms}$ column density (over 10 km s$^{-1}$, the typical velocity width of the \nee\ features) is $1.8 \times 10^{19}\rm~cm^{-2}$. Taking $v_{\rm rel} \sim v_{\rm pec} \approx 19$ km s$^{-1}$ (see Section \ref{sub:environment} and \autoref{fig:localvoid2}), this corresponds to $n_{\rm IGM} \gtrsim 2.3 \times 10^{-4}$ cm$^{-3}$.

Following \citet{benitezllambay+2013}, we also consider the condition that the ram pressure exerted by the IGM, $P_{\rm IGM} \propto n_{\rm IGM} v_{\rm rel}^2$, must overcome the restoring force of the halo, $P_{\rm gal} \propto n_{\rm gas} v_{\rm vir}^2$. We estimate the gas density as the total gas mass of the \nee\ ($M_{\rm gas,\nee} \sim 1.4 M_{\subhi,\rm \nee} = 1.6 \times 10^6$ M$_\odot$) in a sphere of radius $r = 1.9$ kpc (the largest linear size of the \nee; see \autoref{tab:clumps}). We take $v_{\rm vir} = v_{\rm rot} = 17.2$ km s$^{-1}$ for simplicity. This corresponds to $n_{\rm IGM} \gtrsim (v_{\rm vir}/v_{\rm rel})^2 n_{\rm gas} \approx 1.8 \times 10^{-3}$ cm$^{-3}$.

The IGM density depends on a variety of factors, e.g., distance from the galaxy and the temperature of the halo. \citet{mcquinn2016} notes that values typically range from $\Delta_b \sim 5-200$, where $\Delta_b$ is the density in terms of the critical density of the Universe, $n_{\rm crit} \approx 6.2 \times 10^{-6}$ cm$^{-3}$. Assuming $\Delta_b = 10$ (see their Figure 15) yields an IGM density of $n_{\rm IGM} \approx 6.2 \times 10^{-5}$ cm$^{-3}$.

The calculated IGM density and projected morphology of the \nee~allow us to estimate whether ram pressure stripping is at play, but these observations have limitations. Because of the quadratic dependence on the velocity, the threshold IGM density may be significantly different than what we calculate above. Nevertheless, our estimates of the gas density required for ram pressure stripping to be significant do not favor this mechanism as the origin of the \nee. Tangential velocity information is required to determine if the \nee~is pointed opposite the direction of motion (a typical signature of gas stripped by ram pressure) since the line-of-sight velocity alone does not constrain the motion along the direction of the extension. Without proper motion measurements, we cannot constrain the orbit of the galaxy or confirm its 3D motion. However, \autoref{fig:localvoid2} shows that Pisces A is in lockstep with other galaxies near the boundary of the Local Sheet with a nearly vanishing relative peculiar velocity.

Still, ram pressure could have played a significant role if Pisces A originated below the Sheet (at negative $Z_{\rm SG}$) and we are observing the end of its journey through the 3 Mpc thick slab. While this would account for the component of the \nee~direction vector pointed toward negative $Z_{\rm SG}$, it would require a velocity (relative to the Local Sheet) in excess of 1500 km s$^{-1}$ if it also triggered the observed star formation (since the SFH suggests that the formation process began at least 2 Gyr ago). Therefore, while we cannot entirely rule out ram pressure stripping by the IGM, our analysis strongly disfavors it.

\subsection{Past or Present Interactions}\label{sub:interactions}

We must also consider the effects of interactions with other galaxies. It is possible that Pisces A has either experienced, or is presently experiencing, tidal forces due to a host galaxy or a previous encounter, as is seen in a variety of environments \citep[e.g.,][]{smith+2010, pearson+2016, paudel+2018}. The SFH of the galaxy suggests that this would have occured $\gtrsim 2$ Gyr ago, so we should still be able to see some effects of such an encounter, given that the rotational period of Pisces A is $\sim 1$ Gyr. The resulting tidal forces often produce symmetric morphologies. When combined with the morphology of the \nee, the tentatively detected ``southwest extension'' could be evidence of such a tidal encounter.

If tidal forces with a bound host created the \hi\ features we identify, we might expect the required tidal radius, $r_{\rm tid} \sim (M_{\rm dyn, PA} / M_{\rm dyn, host})^{1/3} r_{\rm sep}$, to be on the same order as the furthest extent of the \nee, $\sim 3.1$ kpc (see Table \ref{tab:clumps}). However, this would require a host mass $> 10^{13}$ M$_\odot$ at a separation of 500 kpc, and there is no evidence that Pisces A is bound to such a massive host. In fact, the present-day closest galaxy to Pisces A is AGC 748778, a low-mass ($M_\subhi \sim 3 \times 10^6$ M$_\odot$) dwarf galaxy originally discovered within ALFALFA \citep[][see also \citeauthor{cannon+2011} \citeyear{cannon+2011}]{huang+2012}. At a distance of $\sim 900$ kpc from Pisces A, it is effectively ruled out as the genesis of possible tidal forces.

Pisces A may still have been disrupted at some point in the past by AGC 748778 or another galaxy. One possibility is that Pisces A was once part of an interacting dwarf galaxy pair, and that the companion has since been accreted by a nearby massive host \citep[as observed for other dwarf galaxy pairs by][]{pearson+2016}. The \nee~may therefore have been `pulled' from Pisces A. AGC 748778 is the most obvious candidate for the companion, but its current position does not support such interaction. There may still be another companion that was destroyed or accreted by a massive host on a previous closer passage. A fly by encounter may have injected enough energy to free the gas in the \nee, leaving Pisces A relatively isolated at the present moment. AGC 748778 is the most likely source, with a comparable radial velocity as well as peculiar velocity (relative to the Local Sheet).

A companion galaxy may have also been destroyed by, or combined with, Pisces A via a dwarf-dwarf merger. This scenario has been suggested to trigger recent starbursts \citep[e.g.,][]{noeske+2001, bekki2008}, which would be consistent with the SFH derived by \citet{tollerud+2016}. However, in the case of a dwarf-dwarf merger, the stellar remnant may exist but be too faint to detect.

\citet{nidever+2013} detected a long gas extension ($M_\subhi \sim 7.1 \times 10^5\rm\ M_\odot$) associated with the nearby starburst galaxy IC 10 using deep \hi\ imaging with the Robert C. Byrd Green Bank Telescope. They argue that this feature \citep[and possibly other known \hi\ features, e.g., the ``streamer'' and NE cloud;][]{manthey+2008} is most likely the result of a recent interaction or merger, and that the interaction also triggered a recent ($< 10$ Myr) starburst. They explicitly rule out a stellar feedback origin to the long extension in IC 10 since this would require outflow velocities $\sim 30$ times the observed velocity offset of $-65\rm\ km\ s^{-1}$.

We consider a similar argument for Pisces A. Assuming the furthest observed extent of the anomalous gas (3.1 kpc, measured from \ned), and further assuming the formation of the \nee~began 2 Gyr ago (per the SFH), the required outflow velocities would be on the order of $3.1\rm~kpc / 2~Gyr \approx 1.5\ km\ s^{-1}$. This is in good agreement with the observed velocity offset for \neb~(see Table \ref{tab:clumps}), but fails to explain the other three spatially coincident clumps.

\subsection{Accretion}\label{sub:accretion}

We have so far primarily focused on mechanisms that would \textit{strip} gas from Pisces A, but accretion within the cosmic web also plays an important role in galaxy evolution. In addition to the warm-hot phase \citep[$\sim 10^5 - 10^7$ K; e.g.,][]{evrard+1994, cen+1999} of the IGM, there is also expected to be a cooler \citep[$\lesssim 10^5$ K; e.g.,][]{yoshida+2005} component that can be traced by, amongst other tracers, \hi~emission \citep[][and references therein]{kooistra+2017, kooistra+2019}. ``Cold mode'' accretion of this gas, funneled along cosmic filaments \citep{keres+2005}, can produce tell-tale kinematic signatures in the gas of the accreting galaxy \citep[e.g.,][]{stanonik+2009, kreckel+2011b, ott+2012}, especially for low-mass systems and in low-density environments, such as warped polar disks or misalignment of kinematic axes. Such signatures are notoriously difficult to definitively identify, but it remains a possibility that we are witnessing low-temperature gas falling onto Pisces A via the ``cold mode'' along the \nee. This scenario is supported by the morphology of the \nee, as it is oriented parallel to the plane of the Local Sheet; such accretion activity may then be responsible for the recent star formation implied by the SFH.

\subsection{Pisces A and the Cosmic Web}\label{sub:void}

Although ram pressure is typically invoked as a mechanism for stripping gas \citep[e.g.,][]{benitezllambay+2013}, recent simulations by \citet{wright+2019} show that it may also lead to accretion via compression of halo gas. Using the parallel $N$-body $+$ smoothed particle hydrodynamic code \textsc{gasoline} \citep{wadsley+2004}, they study the evolution of dwarf galaxies ($M_{\rm halo} \sim [0.92 - 8.4] \times 10^9\rm~M_\odot$) that are isolated at $z = 0$. In particular, they follow, in detail, the accretion histories of those galaxies in which star formation shuts off (typically via reionization or supernova feedback) but eventually resumes. These galaxies show significant `gaps' (of at least 2 Gyr but up to 12.8 Gyr) in their SFHs, but have reignited star formation such that almost all are actively forming stars at $z = 0$. They find that `gappy' galaxies have their star formation reignited by encounters with streams of dense gas in the IGM, either associated with the underlying filamentary structure, or produced as the result of mergers of more massive halos also within the cosmic web. \citeauthor{wright+2019} note that the distinguishing feature of these encounters (as compared with dense gas encounters of non-`gappy' galaxies) is the ratio of the ram pressure associated with the gas stream to the galaxy's gravitational restoring force, finding a typical range of $1 \lesssim P_{\rm IGM} / P_{\rm gal} \lesssim 4$. These moderate ram pressure encounters act to \textit{compress}, rather than strip, the gas in the halo onto the disk, allowing the formation of \hi~and H$_2$. The possible motion of Pisces A through a cosmic filament may then require revisiting the ram pressure stripping arguments made in Section 6.1.

The \citet{wright+2019} criteria for defining a `gappy' galaxy is the existence of gaps in the SFH (at $z < 3$) of at least 2 Gyr. By this measure, Pisces A would be considered a `gappy' galaxy (compare Figure 11 of \citeauthor{tollerud+2016} \citeyear{tollerud+2016} with Figure 2 of \citeauthor{wright+2019} \citeyear{wright+2019}). Additionally, all but one of their simulated dwarfs\footnote{The exception is h239c, which they note has a much more chaotic interaction history than the other galaxies in their sample.} has ongoing star formation at $z = 0$ as a result of their dense gas encounters. While \citet{tollerud+2015} are unable to measure an absolute H$\alpha$ flux (and therefore a star formation rate), the presence of any detectable H$\alpha$ emission in Pisces A implies the ongoing, or recent, formation of stars. This suggests that Pisces A has a similar history to the simulated galaxies of \citet{wright+2019}, having a late encounter with dense gas in the IGM within the Local Sheet. In this view, the \nee~could be the result of \hi~gas in the halo of Pisces A being compressed and `falling', or re-accreting, onto the disk.

Pisces A is unlike these `gappy' dwarfs in two distinct ways. First, the simulated galaxies are selected by first simulating the evolution of a massive central halo at a lower resolution, then re-simulating for all galaxy particles that end up within 1 Mpc of the central halo at $z = 0$. This broadly makes them analogs of Local Group galaxies. In contrast, Pisces A does not appear to be associated with any other galaxy at the present epoch. Second, our estimates of gas densities in Section \ref{sub:rampressure} suggest Pisces A is experiencing $P_{\rm IGM} / P_{\rm gal}$ on the order of $10^{-2}$, far too weak for the gas to be compressed. However, \autoref{fig:fbarmhi} suggests that Pisces A is not dissimilar from Local Group dwarfs, and \citet{tollerud+2016} use the size-luminosity relation to show that it may be a candidate for ``prototype'' LG galaxies -- that is, what present-day LG dwarfs may have looked like at earlier epochs. Additionally, the lack of data on the 3D motion of the galaxy, coupled with the heuristic nature of our gas density estimates, means that the ratio of the ram pressure to the galaxy restoring force is poorly understood. More work is needed to constrain the low-temperature IGM in the vicinity of Pisces A. This may be possible with the upcoming Square Kilometer Array, as \citeauthor{kooistra+2017} (\citeyear{kooistra+2017}, \citeyear{kooistra+2019}) show that it is well-suited to provide the first direct detections of \hi~in the IGM.

\section{Conclusions}\label{sec:conclusions}

We present resolved observations of the neutral hydrogen component of Pisces A, while still remaining sensitive to the extended emission surrounding the galaxy, and find an \hi\ mass consistent with previous observations.

With the resolution afforded by the JVLA, we detect multiple morphological and kinematic features in the \hi\ distribution not previously observed. These features are indicative of non-equilibrium, and suggest an active history.

By further quantifying the environment in which Pisces A resides, we confirm its existence within a cosmic filament. We consider a variety of possible origins for the disturbed gas, including ram pressure stripping, past or current galaxy-galaxy interactions, and gas accretion. We find that ongoing interactions or ram pressure stripping are unlikely origins, at least at the present epoch. Our observations provide the strongest support for the idea that the galaxy has recently accreted gas via encounters with streams of gas in the cosmic web, either through the ``cold mode'', ram pressure compression, or some combination of the two.

Because our observations provide only a snapshot of the galaxy, it is still possible one or more of the above mechanisms may have been more significant in the past. The location of Pisces A, and its similarity to the void galaxy population (though with a much lower \hi~fraction and stellar mass than is included in recent surveys of void galaxies), suggests that it may not have originated in the cosmic web. In that case, encounters with gas streams and/or companions would shape its HI morphology during the transition from the void to the web, alternately stripping and accreting gas.

Pisces A provides a unique opportunity to study how the gas content of dwarf galaxies evolves in the context of environment. Future work is required to better understand the IGM near Pisces A, as well as to determine the molecular gas content and metallicity of the anomalous gas clouds, all of which will provide further insights into the processes by which low-mass dwarf galaxies form stars.

\acknowledgments

All land in the United States has been stolen from the indigenous peoples who originally settled it. In particular, the JVLA is located in Socorro, NM, the homeland of various tribes, including the Navajo, Pueblo, and Apache. In Virginia, the NRAO and the University of Virginia occupy land originally settled by a diverse group of nations, especially the Monacan and Mannahoac peoples. It is our responsibility to acknowledge the original stewards of this land.

The authors wish to thank the anonymous referee for comments that improved the quality of this manuscript, and to acknowledge the work of the ALFAFA and GALFA-\hi\ teams. L.B. also wishes to thank A. Towner and J. Hibbard for their invaluable discussions. L. B. was partially supported by the National Radio Astronomy Observatory and the Virginia Space Grant Consortium under grant NNX15A120H.

This research made use of NASA's ADS Bibliographic Services \citep{ads+2004}, as well as the SIMBAD database, operated at CDS, Strasbourg, France \citep{wenger+2000}.

\facility{JVLA}

\software{
CASA \citep{mcmullin+2007},
Astropy \citep{astropy+2018},
APLpy \citep{aplpy+2012},
Matplotlib \citep{matplotlib+2007},
NumPy \citep{numpy+2011},
IPython \citep{ipython+2007},
SciPy \citep{scipy+2001}
}

\bibliographystyle{apj}
\bibliography{references}

\end{document}